\documentclass[pre,groupedaddress,amsmath,amssymb,floatfix,nofootinbib,
balancelastpage,reprint,showpacs,a4paper]{revtex4-1}
\usepackage[pdftex]{graphicx}% Include figure files
\usepackage{amsmath}% Include figure files
\usepackage[usenames]{color}
\usepackage{placeins}
\usepackage{setspace}
\usepackage{bm}

\begin{document}

\title{Rheotaxis of spherical active particles near a planar wall}

\author{W. E. Uspal}
\email[Corresponding author: ]{uspal@is.mpg.de}
\author{M. N. Popescu}
\author{S. Dietrich}
\author{M. Tasinkevych}
\affiliation{Max-Planck-Institut f\"{u}r Intelligente Systeme, Heisenbergstr. 3, D-70569
Stuttgart, Germany}
\affiliation{IV. Institut f\"ur Theoretische Physik, Universit\"{a}t Stuttgart,
Pfaffenwaldring 57, D-70569 Stuttgart, Germany}

\date{\today}

\begin{abstract}

For active particles the interplay between the self-generated hydrodynamic flow and
an external shear flow, especially near bounding surfaces, can result in a rich behavior 
of the particles not easily foreseen from the consideration of the active and external 
driving mechanisms in isolation.  For instance, under certain conditions, the particles 
exhibit ``rheotaxis,'' i.e., they align their direction of motion with the plane of shear spanned by the direction of the flow and the normal of the bounding surface and move 
with or against the flow.  To date, studies of rheotaxis have focused on elongated particles 
(e.g., spermatozoa), for which rheotaxis can be understood intuitively in terms of a ``weather 
vane'' mechanism. Here we investigate the possibility that \textit{spherical} active 
particles, for which the ``weather vane'' mechanism is excluded due to the symmetry of the 
shape, may nevertheless exhibit rheotaxis. Combining analytical and numerical calculations, we 
show that, for a broad class of spherical active particles, rheotactic behavior may 
emerge via a mechanism which involves ``self-trapping'' near a hard wall owing to the 
active propulsion of the particles, combined with their rotation, alignment, and ``locking'' 
of the direction of motion into the shear plane. In this state, the particles move solely 
up- or downstream at a steady height and orientation.

\end{abstract}


\maketitle

\section{Introduction}

Flows are ubiquitous in microconfined fluid systems.  Examples include the flow of blood 
through capillaries, the flow of groundwater through pore spaces in soil, the flow of oil through 
porous rock, and the flow of analyte through the channels of a lab-on-a-chip device. Flow 
presents both challenges and opportunities for natural and engineered active particles, 
especially near surfaces. For instance, fluid shear can trap bacteria near surfaces, inhibiting 
chemotaxis, but promoting surface attachment and biofilm formation \cite{rusconi14}. 
Active particles can harness shear flow for sensing and navigation. A motile 
spermatozoan in shear flow near a surface orients its body axis along the flow direction 
-- a phenomenon known as \textit{rheotaxis}\footnote{In analogy with other ``taxis'' 
phenomena, like chemotaxis, throughout this paper by rheotaxis we shall refer to the alignment 
of a properly defined director axis of the active particle or microorganism along the flow 
direction. Therefore, depending on the relative magnitude of the active motion to that of the 
ambient flow, the net motion of rheotactic particle can yet be downstream even if the 
active component points upstream.} -- and moves upstream \cite{bretherton61}. It is 
thought that in the mammalian oviduct rheotaxis guides sperm to the egg \cite{miki13}. 
Artificial active particles engineered for rheotaxis have promising potential applications.  
For instance, the motion of artificial active particles is generally diffusive over 
sufficiently long timescales because thermal noise eventually randomizes the orientations 
of the particles \cite{howse07,golestanian09}. Rheotactic motion in flow would rectify 
the direction of motion of the particle. Moreover, by analogy with sperm, rheotactic 
active particles could be used for targeted cargo delivery \textit{in vivo} or \textit{in 
vitro}.

Experimental studies of the rheotaxis of sperm near surfaces have a long history. 
Bretherton and Rothschild investigated live and dead sperm under flow in various device 
geometries \cite{bretherton61}. They observed that live sperm always exhibited motion 
upstream, i.e., upstream rheotaxis. 
Subsequently, Rothschild observed that motile sperm tend to accumulate near 
surfaces. He attributed this to an attraction involving the sperm heads 
\cite{rothschild63}.  If heads are attracted to surfaces, then reorientation in flow can be 
qualitatively understood as due to the greater drag on the tail, which points into the bulk, 
than on the head, which is near the surface. With the head acting as a pivot point, the flow 
drags the tail downstream, like a weather vane \cite{winet84,miki13,kantsler14}. Accumulation 
of sperm and other micro-organisms at surfaces is evidently driven by hydrodynamic 
interactions \cite{fauci95,lauga06,berke08,smith09,elgeti10,spagnolie12} and steric repulsion 
\cite{li09,elgeti10,kantsler12,kantsler14}, although the relative significance of these 
contributions is a subject of debate \cite{li09,kantsler12,kantsler14}.

Recently, the experimental efforts have been extended to the study of bacterial rheotaxis. Like 
sperm, bacteria in quiescent fluid are observed to accumulate near surfaces 
\cite{lauga06,berke08,li09}. 
Hill \textit{et al.} studied \textit{E. coli} under flow in a microfluidic channel 
\cite{hill07}. They observed that the bacteria in the middle of the channel tend to 
orient, relative to the flow direction, with a certain acute angle.  They drift to the 
sides of the channel, migrating across fluid streamlines. Upon reaching the side walls, 
the bacteria orient against the flow and swim upstream. As with sperm, these observations 
could be rationalized with a ``weather vane'' mechanism.  More recently, Kaya and 
Koser observed that, within an intermediate band of shear rates, \textit{E. coli} 
can directly align against the flow and swim upstream, without having to migrate to side 
walls \cite{kaya12}.

Furthermore, we note that for micro-organisms in the bulk (far from bounding 
surfaces), the interplay of shear, swimmer geometry, and swimming strategy can drive 
rotation of the body axis towards the vorticity direction (which is perpendicular to the 
shear plane), allowing the organism to migrate across streamlines.  In shear flow, 
bacteria driven by coiled flagella, such as \textit{B. subtillis}, experience a lift force 
in the vorticity direction, owing to their chirality.  Since this chirality is localized 
at the flagella, a torque is exerted on the cell body, driving rotation towards the 
vorticity direction \cite{marcos12}.  Micro-algae also rotate towards the vorticity 
direction.  The mechanism for this orientation is not yet known, but it may be rooted in 
the fact that their two flagella beat at different rates, producing unequal propulsive 
forces and, consequently, a torque on the cell body \cite{chengala13}. This attraction of 
the orientation vector to the vorticity direction has also been labeled rheotaxis 
\cite{marcos12}, but throughout the present study we restrict this notion to denote 
attraction of the orientation vector to the shear plane. 

Theoretical studies have sought to reproduce and isolate generic aspects of swimming in 
flow, including rheotaxis. Z\"{o}ttl and Stark considered a model spherical microswimmer 
driven by Poiseuille flow in narrow tubes and slits \cite{zottl12}. They found that 
hydrodynamic interactions with the confining walls can stabilize upstream swimming.  
So-called ``pullers'' migrated to the centerline and swam upstream, while ``pushers'' 
were attracted to trajectories in which they swim upstream while oscillating between the walls.  
In numerical simulations of elongated run-and-tumble particles (posing as model bacteria) 
driven by Poiseuille flow through slit-like channels, the particles accumulated near walls 
due to steric repulsion between them and the wall and, due to the subsequent alignment of 
their bodies by the flow, swam upstream  \cite{nash10,costanzo12}. 
Chilukuri \textit{et al.} modeled swimmers as Brownian ``dumb-bells'' 
(i.e., two beads connected by a spring) \cite{chilukuri14}.  Likewise, when 
driven by Poiseuille flow through a wide slit, a fraction of the particles accumulated 
near surfaces and swam upstream. This accumulation was enhanced upon including 
hydrodynamic interactions between the dumb-bells and the walls, indicating that, for 
this system, both steric repulsion and hydrodynamic interactions promote rheotaxis. 
Additionally, for a Brownian self-propelled particle (i.e., one exposed to stochastic 
forces), ten Hagen \textit{et al.} found that an unbounded shear flow can modify the scaling with time of the mean squared displacement of the particle \cite{tenhagen11}.

In the last decade, significant progress has been made in developing synthetic 
self-propelled particles. These particles are envisioned as key components in future lab-on-a-chip, drug delivery, and pollution 
remediation systems \cite{ebbens10, patra12}. In these and other applications, active 
particles will undoubtedly encounter shear flows near surfaces and will need to 
autonomously sense and respond to flow. Many generic aspects of artificial 
microswimmers in flow are likely captured by the theoretical and numerical 
studies discussed above. However, achieving rheotaxis in artificial systems, for which active stresses can be generated through, e.g., local changes in the chemistry of the 
environment, requires an in-depth analysis of the interplay between the external 
flow, the presence of bounding surfaces, and the mechanism of propulsion 
(i.e., activity). For example, for a large class of active particles, a reaction 
involving a ``fuel'' molecule in the surrounding solution is catalyzed by a region of the 
particle surface, leading to propulsion either via the formation and expulsion of 
bubbles \cite{sanchez11}, or via ``self-phoresis,'' i.e., the generation of a 
tangential surface flow powered by the gradient of the product and/or reactant 
molecules \cite{anderson89,golestanian05}. For the latter case, we have recently shown that the 
complex behavior emerging when the particle motion occurs near confining boundaries 
can be understood only if the coupling loop between the distributions of chemical species 
and hydrodynamics is fully accounted for \cite{uspal14}.

There have been comparatively few studies of artificial microswimmers in flow, and the 
issue of rheotactic behavior was rarely raised. Sanchez \textit{et al.} have 
shown that bubble-powered, rod-shaped catalytic ``microjets'' can move upstream in a 
channel when guided by an external magnetic field \cite{sanchez11}. 
Frankel and Khair studied theoretically a model of self-phoretic Janus particles and 
have shown that drift across streamlines in unbounded shear flow may occur at finite 
P\'{e}clet number \cite{frankel14}. Tao and Kapral have used numerical simulations 
in order to 
investigate the motion of a self-phoretic dimeric nanomotor moving upstream against 
a Poiseuille flow in a channel and to examine the effect of confined flow on the 
propulsion velocity of the motor \cite{tao10}. However, they used a potential which 
confined the motor to the channel centerline, and therefore did not probe the 
possibility of rheotaxis. Recently, Palacci 
\textit{et al.} employed microfluidic chips and self-propelled dimers made out of 
polymer spheres and hematite cubes partially embedded in the polymer matrix in order to provide 
the first experimental demonstration of rheotaxis, via the ``weather vane'' mechanism, for 
an artificial swimmer \cite{palacci14}. The heavy dimers sediment to the bottom surface 
and, once being exposed to blue light and to hydrogen peroxide 
``fuel'', the hematite becomes catalytically active and promotes the 
decomposition of the hydrogen peroxide. This leads to an attraction of the hematite end to the surface and thus, while the whole dimer tends to move due to the 
concentration gradients, the hematite end serves as an ``anchor'' for the dimer to 
orient with the ambient, externally controlled flow within the microfluidic chip.  
\begin{figure*}[!htb]
\includegraphics[width=0.8\textwidth]{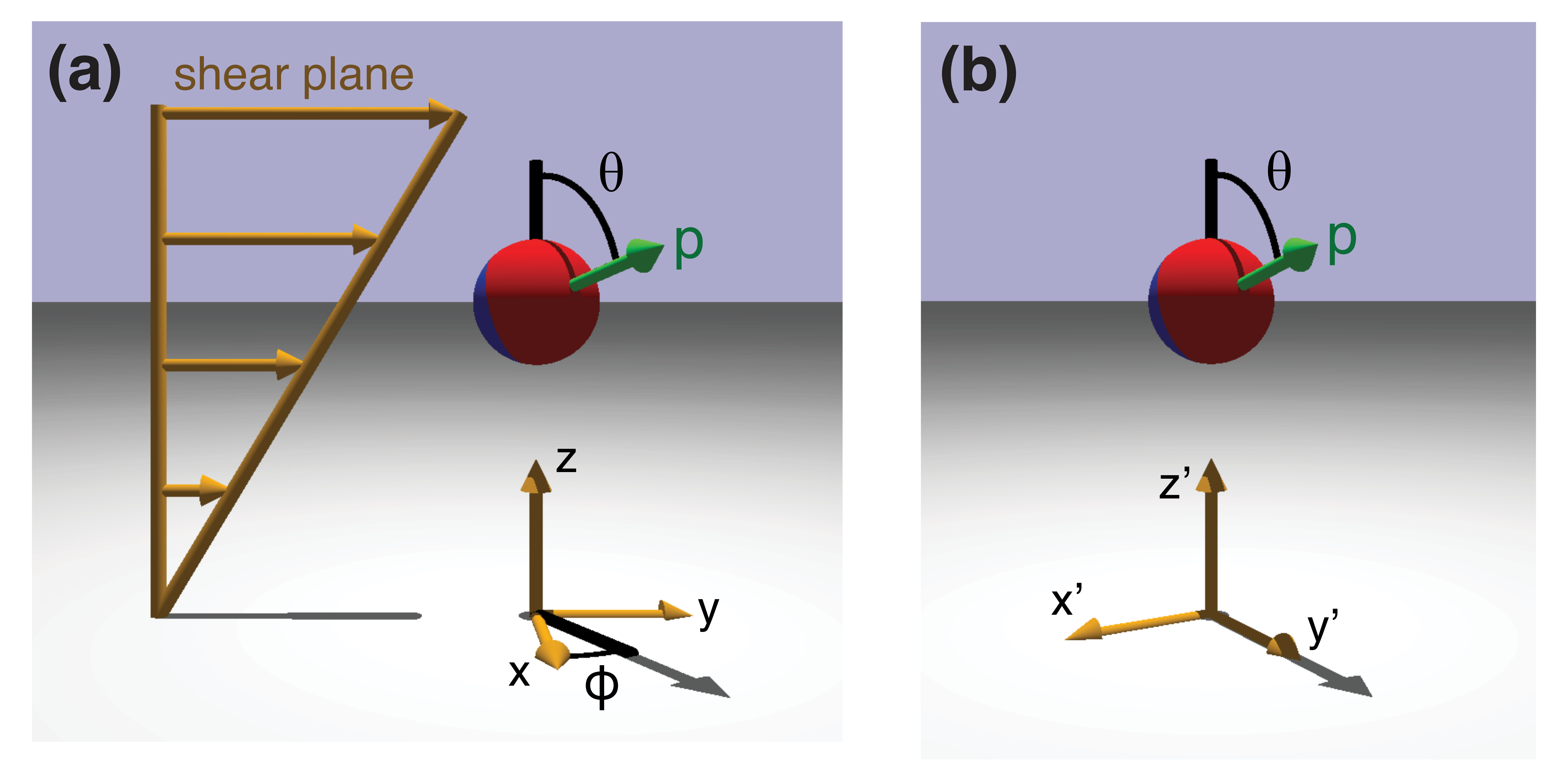}
\caption{\label{fig:schematic_3d} (a) A spherical active particle is suspended in a 
linear shear flow above a planar wall which is translationally invariant in $x$ 
and $y$ directions.  Here, it is shown as a Janus particle which moves away 
from its catalytically active cap (blue).  The orientation of the particle is 
indicated by the unit vector $\mathbf{p}$ (green), which is described by the 
angles $\theta$ and $\phi$. Black arrows and lines correspond to projections onto 
the $xy$ and $x'y'$ plane. (b) Illustration of the ``primed'' coordinate frame used to 
solve subproblem II (see main text). In both (a) and (b) the planar wall is 
located at $z = 0$ and $z' = 0$, respectively.
}
\end{figure*}

In the present study we address the issue of whether artificial \textit{spherical} 
active particles can exhibit rheotaxis when moving in the vicinity of a planar 
surface. Without a distinction between major and minor axes, the particle cannot align 
via 
the intuitive ``weather vane'' effect. Thus it is not \textit{a priori} clear if 
rheotaxis 
is possible for such systems; for rheotatic behavior to emerge nonetheless a 
different physical mechanism must be at work. Considering the overwhelming importance of 
spherical particles for experiments and applications of artificial swimmers (a 
spherical particle is naturally simpler and easier to fabricate than an elongated body), 
this question out of basic research turns out to be one of significant 
practical interest, too. Furthermore, the answer to this question may also shed light on 
the motility in confined flows of micro-organisms with spherical or nearly 
spherical shape.

We focus our theoretical and numerical study on the case of spherical particles with an 
axisymmetric propulsion mechanism which can be described via an effective surface 
slip velocity. This framework captures a number of important classes of microswimmers, 
including self-phoretic catalytically active Janus particles, as well as the classical 
``squirmer'' model, introduced by Lighthill \cite{lighthill52} and subsequently refined 
by 
Blake \cite{blake71}, which is known to  account very well for the essential features of 
the self-propulsion of ciliated micro-organisms. We show that if the swimming 
activity of the particle can lead to a ``self-trapping'' near a bounding surface 
rheotactic behavior may emerge. We establish the conditions under which this 
``self-trapping'' is accompanied by a mechanism for resisting rotation by the flow and 
maintaining a steady orientation. In contrast with the ``weather vane'' mechanism, 
rotation of the orientation vector (i.e., the symmetry axis of the particle) into the 
shear plane is now driven not by the external flow, but rather by the near-surface 
swimming activity of the particle. Building on these results, as well as on our recent 
study of the self-propulsion of a catalytic Janus particle near a surface in a 
quiescent fluid \cite{uspal14}, we show how the surface chemistry of a catalytically 
active Janus particle can be designed to achieve rheotaxis. The wide applicability of our 
results is further emphasized by demonstrating that rheotactic behavior also occurs for 
squirmers which in quiescent fluids are attracted by surfaces \cite{ishimoto13,li14}.

\section{\label{theory}Theory}

\subsection{Problem formulation}

We consider a spherical, axisymmetric, active particle suspended in a linear shear 
flow in the vicinity of a planar wall (Fig. \ref{fig:schematic_3d}(a)).  We adopt a 
coordinate frame (the ``lab frame'') in which the wall is stationary.  The wall is 
located 
at $z = 0$ and has a normal vector in the $\hat{z}$ direction. The particle has radius 
$R$ 
and its center is located at $\mathbf{r}_{p} = (x_{p}, y_{p}, z_{p})^{T}$; 
$h = z_{p}$ denotes the height of the particle center above the wall. We 
describe the particle orientation in terms of the ``director'' $\mathbf{p}$, 
$|\mathbf{p}|=1$, which is the direction of active motion if the particle would be 
suspended in an unconfined, quiescent fluid. Due to axial symmetry, $\mathbf{p}$ 
is parallel to the axis of symmetry of the particle. (For instance, for a catalytic Janus 
particle which moves ``away'' from its catalytic cap, the direction of $\mathbf{p}$ is 
given by the vector from the pole of the catalytic cap to the pole of the inert region.) 
The height and orientation completely specify the instantaneous particle configuration. 
We 
shall find it convenient to alternatively describe the particle orientation 
in terms of spherical coordinates $\theta$ and $\phi$, where $\theta$ is the angle 
between $\mathbf{p}$ and $\hat{z}$, and $\phi$ is the angle between $\hat{x}$ and the 
projection of $\mathbf{p }$ onto the $xy$ plane (see Fig. \ref{fig:schematic_3d}(a)). We 
seek to calculate the translational velocity $\mathbf{U}$ and angular velocity 
$\mathbf{\Omega}$ of the particle as a function of the configuration ($h$, $\mathbf{p}$).

The active particles we consider here are such that the self-propulsion 
mechanism can be accounted for via an effective surface slip velocity $\mathbf{v}_{s}$. 
The slip velocity can encode, for example, the surface flow generated by a ``squirming'' 
mechanical deformation of the particle surface (e.g., the time averaged motion 
of the cilia of a micro-organism), or the surface flow generated by the interactions 
between reactant and product molecules with the particle (as in the case of a 
catalytically active Janus colloid). Here we take the slip velocity as a known function 
of the position along the particle surface and of the configuration ($h$, $\mathbf{p}$). 
We postpone for Section II.E the discussion of how to obtain this slip velocity 
for a given swimmer model.  

The velocity in the suspending fluid is $\mathbf{u}(\mathbf{r}) = 
\mathbf{u}_{ext}(\mathbf{r}) + \mathbf{u}_d(\mathbf{r})$, where 
$\mathbf{u}_{ext}(\mathbf{r}) = \dot{\gamma} z \hat{y}$ is the \textit{ext}ernal 
background flow, $\mathbf{u}_d$ is the \textit{d}isturbance flow created by the 
particle, and $\dot{\gamma}$ denotes the spatially and temporally constant shear 
rate. The fluid velocity is governed by the Stokes equation $-\nabla P + \eta \nabla^{2} 
\mathbf{u} = 0$, where $P(\mathbf{r})$ is the pressure field and $\eta$ is the viscosity 
of the solution, as well as the incompressibility condition $\nabla \cdot \mathbf{u} = 
0$. 
On the planar wall, the velocity satisfies the no-slip boundary condition 
$\mathbf{u} = 0$.  On the surface of the particle, accounting for the effective slip 
implies that the fluid velocity satisfies $\mathbf{u} = \mathbf{U}  + \mathbf{\Omega} 
\times (\mathbf{r} - \mathbf{r}_{p}) + \mathbf{v}_{s}$. Finally, we specify that the 
active particle is force and torque free, thus closing the system of equations 
for the six unknowns $\mathbf{U}$ and $\mathbf{\Omega}$.

\subsection{Solution by linear superposition}

We exploit the linearity of the Stokes equation in order to split the complete problem into two 
subproblems.  In subproblem I we consider the motion of an inert (non-active) 
sphere with configuration $(h,\,\mathbf{p})$ driven by an external shear flow. We note 
that although, for consistency of notations, $\mathbf{p}$ is included as a configuration 
variable, in the case considered in this subproblem the velocity of the particle actually 
does not depend on $\mathbf{p}$ (but, obviously, $\mathbf{p}$ will evolve in time due to 
rotation of the particle.) In subproblem II, a spherical 
active particle with instantaneous configuration $(h,\,\mathbf{p})$ moves through a 
quiescent fluid. The particle velocity for the complete problem is then obtained by 
linear superposition: $\mathbf{U} = \mathbf{U}_{f} + \mathbf{U}_{a}$ and $\mathbf{\Omega} 
= \mathbf{\Omega}_{f} + \mathbf{\Omega}_{a}$, where the velocities $\mathbf{U}_{f}$ and 
$\mathbf{\Omega}_{f}$ are due to \textit{f}low (subproblem I), and the velocities  
$\mathbf{U}_{a}$ and $\mathbf{\Omega}_{a}$ are due to the \textit{a}ctivity-induced motion (subproblem II).

Subproblem I was solved analytically by Goldman \textit{et al.} using bispherical 
coordinates \cite{goldman67}. The translational and rotational velocities of the particle 
are given by
\begin{eqnarray}
\label{sol_subprob_I}
\mathbf{U}_{f} &=& \dot{\gamma} h g(h/R) \hat{y}\,,\nonumber\\
\mathbf{\Omega}_{f} &=&  -\frac{1}{2} \dot{\gamma} f(h/R) \hat{x}\,.
\end{eqnarray}
Note that the external flow does not cause motion of the particle in the $\hat{z}$ 
direction. The functions $f(h/R)$ and $g(h/R)$ encode the hydrodynamic interaction with 
the no-slip wall, which retards both the translation and the rotation of the particle. 
These functions increase monotonically with $h/R$ and have the properties $0 < f(h/R) < 
1$, $0 < g(h/R) < 1$, $f(h/R \to \infty) \rightarrow 1$, and $g(h/R \to \infty) 
\rightarrow 1$. They are tabulated for selected values of $h/R$ in Ref. 
\cite{goldman67}. 

We now turn to subproblem II.  Due to the symmetry of the system and the 
absence of thermal fluctuations, the motion of the particle is confined to the plane 
spanned by the orientation vector $\mathbf{p}$ and the wall normal $\hat{z}$.  
Therefore it is convenient to introduce a ``primed'' reference frame to solve this 
subproblem (Fig. \ref{fig:schematic_3d}(b)).  In the primed frame, the wall is stationary 
and located at $z' = 0$.  The vector $\hat{z}'$ is normal to the wall, the 
vector $\hat{y}'$ lies in the plane containing the particle symmetry axis and $\hat{z}'$ 
(which means that the projection of $\mathbf{p}$ onto the $x'y'$ plane points into 
the direction of $\hat{y}'$), and $\hat{x}'$ is orthogonal to $\hat{y}'$ and $\hat{z}'$. In 
general, this subproblem must be solved numerically. We discuss the numerical 
solution in Section III.  Here we assume the solution to be known and we use the most 
general form compatible with the symmetry constraints discussed above, i.e., a 
translational velocity $\mathbf{U}_{a}' = U_{a,y'} \hat{y}' + U_{a,z'} \hat{z}'$ 
and an angular velocity $\mathbf{\Omega}_{a}' = \Omega_{a,x'} \hat{x}'$, where 
$U_{a,y'}$, $U_{a,z'}$, and $\Omega_{a,x'}$ are functions of 
$h/R$ and $\theta$. Transforming back to the original reference frame, we obtain
\begin{eqnarray}
\label{Ua_def}
\mathbf{U}_{a} &=& U_{a,y'} \cos(\phi) \hat{x} + U_{a,y'} \sin(\phi) \hat{y} + 
U_{a,z'} \hat{z}\,,
\nonumber\\
\mathbf{\Omega}_{a} &=&  \Omega_{a,x'} \sin(\phi) \hat{x} - \Omega_{a,x'}  
\cos(\phi) \hat{y}\,.
\end{eqnarray}

\subsection{Particle motion and fixed points for the dynamics}

In the absence of thermal fluctuations (which we assume to be negligible), the time 
evolution of the particle configuration $(h,\,\mathbf{p}$) is determined, in the 
overdamped limit, by the translational and rotational velocities derived above. By noting 
that the motion of the director obeys $\dot{\mathbf{p}} = \mathbf{\Omega} \times 
\mathbf{p}$ and that (see Fig. \ref{fig:schematic_3d}(a))
\begin{equation}
 \sin(\phi) = \dfrac{p_y}{\sqrt{1-p_z^2}}\,,~~\cos(\phi) = 
\dfrac{p_x}{\sqrt{1-p_z^2}}\,,\nonumber
\end{equation}
we arrive at the following dynamical equations (using $\mathbf{\Omega} = 
\mathbf{\Omega}_f + \mathbf{\Omega}_a$, $\mathbf{U} = \mathbf{U}_f + \mathbf{U}_a$, 
$\dot{\mathbf{p}} = \mathbf{\Omega} \times \mathbf{p}$):
\begin{eqnarray}
\label{eq:dynamical_eqns}
\dot{p}_{x} &=& \Omega_y p_z - \Omega_z p_y =   -\frac{\Omega_{a,x'}(p_{z}, 
h/R) p_{x} p_{z}}{\sqrt{1 - p_{z}^{2}}}\,, \label{dot_px} \\
\dot{p}_{y} &=&  \Omega_z p_x - \Omega_x p_z \nonumber \\
&=&  \, \frac{1}{2} \dot{\gamma} p_{z} 
f(h/R) - \frac{\Omega_{a,x'}(p_{z}, h/R) p_{y} p_{z}}{\sqrt{1 - p_{z}^{2}}} \,,
\label{dot_py} \\
\dot{p}_{z} &=& \Omega_x p_y - \Omega_y p_x \nonumber \\
&=&   -\frac{1}{2} \dot{\gamma} p_{y} 
f(h/R) + \Omega_{a,x'}(p_{z}, h/R) \sqrt{1 - p_{z}^2} \,, \label{dot_pz} \\
&& \hspace*{-0.65in} \mathrm{and}\nonumber\\
\dot{h}_{~} &=& U_z = \, U_{a,z'}(p_{z}, h/R)\,, \label{dot_h}
\end{eqnarray}
where the dot over a quantity denotes its time derivative. Since $p_{x}^{2} + p_{y}^{2} + p_{z}^{2} = 1$, the dynamics of the director must 
obey the constraint  $p_{x} \dot{p}_{x} + p_{y} \dot{p}_{y} + p_{z} \dot{p}_{z} \equiv 0$; 
this is obviously satisfied by Eqs. (\ref{dot_px})-(\ref{dot_pz}). Since, 
accordingly, the three components of $\mathbf{p}$ are not independent, the dynamical 
system is \textit{de facto} three-dimensional.

Defining the generalized velocity vector $\mathbf{V} \equiv [\dot{p}_{x}, \dot{p}_{y}, 
\dot{p}_{z}, \dot{h}]^{T}$, we search for fixed points $(\mathbf{p}^{*}, h^{*})$ of the 
dynamics, which are determined by $\mathbf{V}(\mathbf{p}^{*}, h^{*}) = 0$. Such 
configurations $(\mathbf{p}^{*}, h^{*})$ correspond to the particle translating along the 
wall with a fixed height and orientation, i.e., steady states with properties 
which are compatible with (although not all of them are necessary for) rheotaxis. 
In order to obtain $(\mathbf{p}^{*}, h^{*})$, we start by considering the three 
possibilities for $\dot{p}_{x} = 0$ to be satisfied. For completeness, we discuss here all 
three cases and, for each of them, derive the additional conditions following from 
$\dot{p}_{y} = 0$, $\dot{p}_{z} = 0$, and $\dot{h} = 0$. In anticipation, we note that for 
the types of swimmer considered here, it turns out that only the third case, which 
corresponds to $p_{x}^{*} = 0$, is compatible with rheotaxis. Briefly, in the first two cases, the combined effects of shear and activity produce a fixed point with $p_{z}^{*} = 0$, i.e., $\theta = 90^{\circ}$. Most types of swimmer do not satisfy the restrictive conditions required to produce such a fixed point. Consequently, the rest of the study in this paper will focus on the third case.

(i) The case $\Omega_{a,x'}(p_z = \, p_{z}^{*}, h^{*}) = 0$. This is a particle 
configuration for which in the absence of external flow the particle only translates. In 
order to also satisfy $\dot{p}_{y} = 0$, $\dot{p}_{z} = 0$, and $\dot{h} = 0$, it is 
then required that $p_{y}^{*} = 0$, $p_{z}^{*} = 0$, and $U_{a,z'}(p_{z}^{*} = 0, 
h^{*}) = 0$. The first two of the latter relations imply that the director is along 
the $\hat{x}$-axis, which coincides with the vorticity axis of the external shear 
flow\footnote{The vorticity $\boldsymbol{\omega}$ of the flow 
$\mathbf{u}$ is defined as $\boldsymbol{\omega} = \nabla \times \mathbf{u}$; for the 
shear flow $\mathbf{u}_{ext} = \dot{\gamma} z \hat y$ one has $\boldsymbol{\omega} = -\dot{\gamma} \hat x$ so that the vorticity direction is thus everywhere aligned with the 
$x$-axis.}; thus the director is not rotated by the external flow, while the entire 
particle rotates around the director due to the external shear flow, which 
also advects the particle (a ``log rolling'' state). Furthermore, with $p_{z}^{*} = 0$ 
the conditions $\Omega_{a}'(p_{z}^{*} = 0, h^{*}) = 0$ and $U_{a,z'}(p_{z}^{*} = 0, 
h^{*}) = 0$ imply that there must be a fixed point for subproblem II which occurs at 
$p_{z}^{*} = 0$, i.e., at $\theta^{*} = 90^{\circ}$, for which the director is 
parallel to the wall (see Fig. \ref{fig:schematic_3d}(a)). We note that this condition 
on subproblem II is very restrictive; it turns out that it cannot be satisfied by 
any of the active particles considered in Sec. III.

(ii) The case $p_{z}^{*} = 0$. In this case the director lies within a plane 
parallel to the wall (see Fig. \ref{fig:schematic_3d}(a)). The equation $\dot{p}_{y}= 0$ 
is then automatically satisfied (Eq. (\ref{dot_py})).  The condition $\dot{h} = 0$ 
implies that the curve $p_z(h)$ defined by $U_{a,z'}(p_{z}, h) = 0$ 
passes through $\theta = 90^{\circ}$ (i.e., $p_z= 0$) at a certain height $h^{*}$. 
The condition $\dot{p}_{z} = 0$ then implies $p_{y}^{*} = \frac{2 
\Omega_{a,x'}(p_{z}^{*} = 0, h^{*})}{\dot{\gamma} f(h^{*}/R)}$. The corresponding 
physical picture is as follows. As the director is within a plane parallel to the wall, 
the contributions to $\dot{\mathbf{p}}$ from both shear and activity are entirely along 
the $\hat{z}$ direction: for $p_z = 0$ Eqs. (\ref{dot_px}) - (\ref{dot_pz}) imply 
$\dot{\mathbf{p}} = (0,0,\dot{p_z})$. The magnitude of the contribution from the shear 
flow ($\sim \dot{\gamma}$ in Eq. (\ref{dot_pz})) depends on the angle $\phi$.  For 
instance, if the director is oriented along the $x$-axis ($p_y = 0$, so that 
$\phi = 0,\,\pi$), the contribution of shear to $\dot{p}_{z}$ is zero (see Eq. 
(\ref{dot_pz})).  At $p_y = p_{y}^{*}$ (see above, i.e., for a certain angle $\phi 
\neq 0,\pi$), due to the definition of $p_y^*$ the contributions from shear and 
activity to $\dot{p}_{z}$ sum to zero. For this fixed point to occur, due to 
$|\mathbf{p}| = 1$ it is required that $|p_y^*| = \left| \frac{2 
\Omega_{a,x'}(p_{z}^{*} = 0, h^{*})}{\dot{\gamma} f(h^{*}/R)} \right| \leq 1$.  This 
requirement, together with that of the vanishing of $U_{a,z'}$ at a certain height 
$h^{*}$ for $\theta = 90^{\circ}$, places weaker demands on subproblem II than the 
requirements in case (i). Nevertheless, as in case (i), it turns out that none of the 
types of active particles which will be considered in Sec. III satisfies the fixed point 
requirement on $U_{a,z'}$, and thus none of them can fit case (ii) either. 

%%%%%%%%%%%%%%%%%%%%%%%%%%%%%%%%%%
\textbf(iii) The case $p_{x}^{*} = 0$. In this case the director $\mathbf{p}$ 
lies within the shear plane (i.e., the $yz$ plane in Fig. 
\ref{fig:schematic_3d}(a)). If a fixed point exists, in that state the particle 
translates either \textit{only} upstream or \textit{only} downstream because once 
having reached the fixed point of the dynamics the orientation cannot vary in 
time. If the fixed point is stable (this issue will be addressed analytically in 
the next subsection and numerically in Sec. III), this state corresponds to 
\textit{rheotaxis}.  Since in this case $p_{y} = 
\pm \sqrt{1 - p_{z}^{2}}$ and thus $p_y \,\dot{p}_y = - p_z \,\dot{p}_z$, 
Eqs. (\ref{dot_py}) and (\ref{dot_pz}) are identical (as expected, because only 
two of the Eqs. (\ref{dot_px}) - (\ref{dot_pz}) are independent). Therefore, the steady height 
$h^{*}$ and orientation $p_{z}^{*}$ are obtained as the solutions of 
\begin{eqnarray}
\label{eq:rheotactic_state}
&&\mathrm{sgn}(p_{y}^{*}) \, \Omega_{a,x'}(p_{z}^{*}, h^{*}/R) = \frac{1}{2} 
\dot{\gamma} f(h^{*}/R)\,,
\nonumber\\
&& U_{a,z'}(p_{z}^{*}, h^{*}/R) = 0\,.
\end{eqnarray}
The first equation simply expresses that at the steady state the sum of the contributions 
to the angular velocities from shear and activity must add up to zero. The term 
$\text{sgn}(p_{y}^{*})$ appears because shear tends to rotate the director away from the 
wall when $p_{y} < 0$ (i.e., $\phi = 270^{\circ}$) and towards the wall when $p_{y} > 0$ 
(i.e., $\phi = 90^{\circ}$) (see Fig. \ref{fig:schematic_3d}(a) and recall 
that in the present case $\mathbf{p}$ lies in the $yz$ plane), whereas, owing to the 
symmetry of subproblem II, there is no dependence on $\phi$ for the direction of 
rotation by the swimming activity. In contrast to the cases (i) and (ii), it turns out 
that the conditions in Eqs. (\ref{eq:rheotactic_state}) can be satisfied by all three 
classes of spherical active particles considered here (see Sec. III). 
Therefore we proceed further with the analysis of the case $p_x^* = 0$.
%%%%%%%%%%%%%%%%%%%%%%%%%%%%%%%%%%
\begin{figure*}[!htp]
\includegraphics[width=0.95\textwidth]{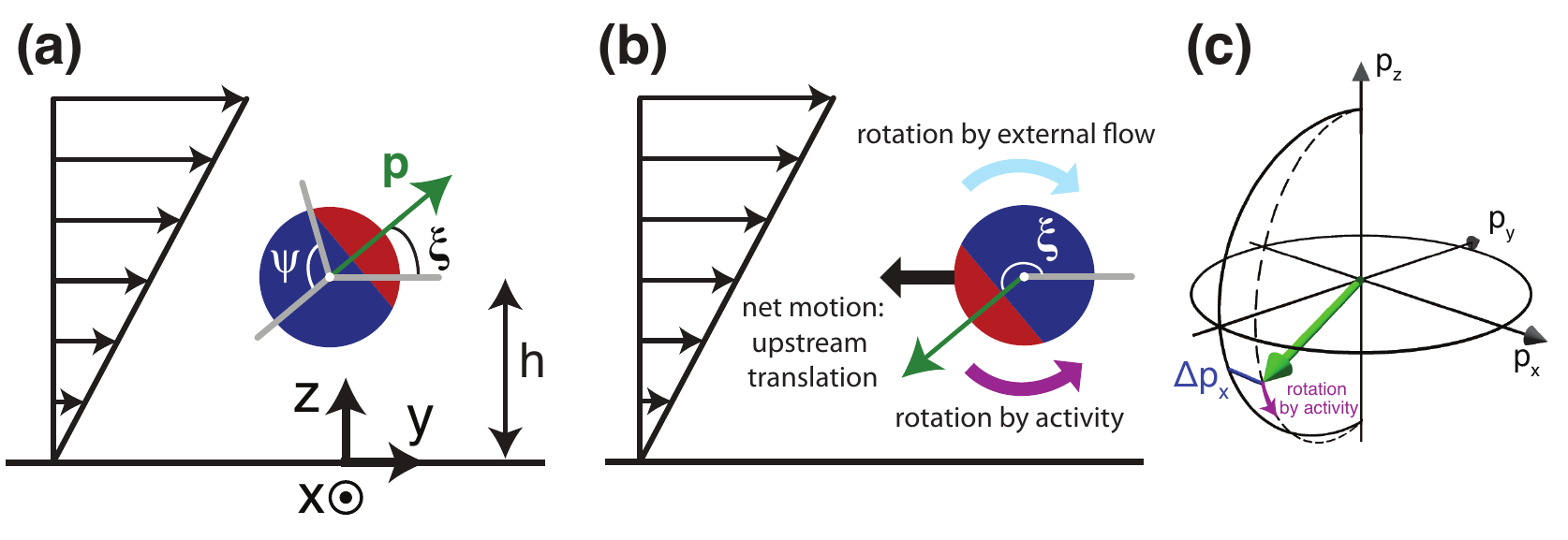}
\caption{
\label{fig:xi_rheotaxis_schematic} 
(a) Schematic description of the particle dynamics if the director 
$\mathbf{p} = (p_x,p_y,p_z)$ (green) lies in the indicated shear plane.  
The angle $\xi$ is formed by the director and the flow direction. The wall is 
located at $z=0$ and $h$ is the distance of the particle center from the wall. For a 
catalytic Janus particle, the angle $\psi$ characterizes the size of the catalytic cap 
(blue). Note that here $\psi$ is larger than in Fig. \ref{fig:schematic_3d}. (b) 
Illustration of a particle in a rheotactic steady state which occurs for $p_{y}^{*} < 0$ 
and $p_{z}^{*} < 0$, i.e., for $\xi$ within the range  $\pi < \xi < 3 \pi/2$. The 
external flow contributes a clockwise component $-\frac{1}{2} \dot{\gamma} f(h/R) 
\hat{x}$ 
(cyan) to the particle angular velocity (Eq. (\ref{sol_subprob_I})).  For the angle $\xi$ 
to remain stationary, the swimming activity of the particle must contribute a 
counterclockwise component (magenta) with equal magnitude to the angular velocity.  Since 
the particle, assumed to be axisymmetric, does not rotate in free space, i.e., 
without the wall and in the absence of the external shear flow, this contribution must 
be due to the interaction with the wall. If the swimming velocity of the particle is 
large enough, the particle swims upstream. (c) Upstream rheotaxis is stable against 
perturbations of the director out of the shear plane, where $p_x= 0$. Shear 
flow tends to rotate the director around the $\hat{x}$ direction, keeping $p_x$ 
constant and thus does not contribute to $\dot{p}_{x}$ (see the last 
paragraph in Subsec. II.C). The swimming activity of the particle tends to rotate the 
orientation vector along ``lines of longitude'' on the unit sphere $|\mathbf{p}| = 1$. 
For a small perturbation $\Delta p_{x}$ (blue line segment) away from the rheotactic 
state 
shown in (b), activity rotates the director (green vector) towards the pole $p_{z} 
= -1$, decreasing $p_{x}$ and $p_y$ due to $|\mathbf{p}| = 1$. The effect of activity is 
indicated by the magenta arrow.
}
\end{figure*}

\subsection{Linear stability analysis of the fixed point $p_{x}^{*} = 0$ 
\label{section:linear_stability}}

In order to establish the conditions under which the fixed point $p_{x}^{*} = 0$ (case 
(iii) above) is stable, so that the corresponding steady state exhibits rheotaxis, we proceed by carrying out a linear stability analysis. Since for 
$p_{x} = 0$ the director lies within the shear plane and, due to the symmetry of the 
problem, cannot rotate out of this plane, in the three-dimensional phase space defined by 
$p_{x}$, $p_{z}$, and $h$ a trajectory with initial condition in the plane $p_{x} = 0$ 
will be confined to that plane for all times $t$.  Upon linearization of the equations 
of motion near the fixed point, a small perturbation $\Delta p_{x}$ 
decouples from the other variables:
\begin{equation}
\frac{d \Delta {p}_{x}}{dt} = -\frac{\Omega_{a,x'}(p_{z}^{*}, h^{*}/R) 
p_{z}^{*}}{\sqrt{1 - p_{z}^{*2}}} \Delta {p}_{x}.
\end{equation}
After using Eq. (\ref{eq:rheotactic_state}) for substituting\newline
$\text{sgn}(p_{y}^{*}) \, \Omega_{a,x'}(p_{z}^{*}, h^{*}/R) = \frac{1}{2} 
\dot{\gamma} 
f(h^{*}/R)$, we obtain
\begin{equation}
\frac{d \Delta {p}_{x}}{dt} = -\frac{\dot{\gamma} f(h^{*}/R) p_{z}^{*}}{2 p_{y}^{*}} 
\Delta {p}_{x}\,.
\end{equation}
Therefore $\Delta p_{x}$ decays or grows exponentially,
\begin{equation}
\label{eq:px_exp_decay}
\Delta {p}_{x} = 
\Delta p_{x}(0) \exp\left( \frac{-\dot{\gamma} f(h^{*}/R) p_{z}^{*}}{2 p_{y}^{*}}\, t
\right)\,,
\end{equation}
depending on the sign of the quantity multiplying $t$ in the argument of the 
exponential.

Since $f(h^{*}/R) > 0$ and $\dot{\gamma} > 0$, the stability of the fixed point is 
determined by the sign of the ratio $p_{z}^{*}/p_{y}^{*}$. For the particle to swim 
oriented upstream (i.e., to exhibit ``positive'' rheotaxis), it must align against the 
flow, so that $p_{y}^{*} < 0$; stability then requires that $p_{z}^{*} < 0$, i.e., the 
director points towards the wall.  Additionally, if the swimming velocity is large enough 
to overcome the shear flow, i.e., for $\dot{y}_{p} < 0$, where $\dot{y}_{p} = 
-U_{a,y'} + \dot{\gamma} h^{*} g(h^{*}/R)$ (see $U_{f,y}$ in Eq. 
(\ref{sol_subprob_I}) and $U_{a,y}$ in Eq. (\ref{Ua_def}) with $\phi = 
270^\circ$), \textit{net} upstream swimming emerges. The other possibility, that an 
attractor exists for which $p_{y}^{*}$ and $p_{z}^{*}$ are both positive, is in general 
unlikely, 
because the particle would be pointing away from the wall (see Fig. 
\ref{fig:schematic_3d}(a)). 
Furthermore, it would correspond to the less interesting case of \textit{negative} 
rheotaxis, 
in which the particle motion is in the flow direction. Therefore, it will not be discussed further.

The analysis of the linear stability shows that we need to examine particle 
dynamics only in the symmetry plane $p_{x} = 0$ in order to determine whether rheotaxis occurs. 
In the following, we shall examine the two-dimensional dynamics in this plane in detail. 
Additionally, we shall perform fully three-dimensional numerical calculations of 
trajectories with initial 
conditions $p_{x}(t = 0) \neq 0$ in order to explore the limits of validity of 
the linear analysis, as well as to understand the details of the evolution towards 
alignment in the shear plane.  We introduce a new angle $\xi$, shown in Fig. 
\ref{fig:xi_rheotaxis_schematic}(a), which will turn out to be convenient for 
describing the director orientation if $p_{x} = 0$.  The requirement that the 
attractor occurs with $p_{y}^{*} < 0$ and $p_{z}^{*} < 0$ translates into $\pi < 
\xi^{*} < 3 \pi/2$.

Before proceeding with the analysis, we discuss the physical picture behind the stability 
criteria in Eq. (\ref{eq:px_exp_decay}). While we have provided mathematical conditions 
for rheotaxis, the corresponding physical interpretation can provide further insight. 
Equation (\ref{eq:dynamical_eqns}) shows that particle activity has an effect on $p_{x}$, but 
shear does not. The contribution of shear to the angular velocity of the particle is 
rigid rotation around the $\hat{x}$ axis.  In the absence of activity, shear rotates the 
tip of the director in a coaxial circle around the $\hat{x}$ axis, keeping 
$p_x$ constant.  Now we consider activity. Due to the mirror symmetry with respect to the plane containing $\mathbf{p}$, the contribution 
of activity can only drive rotation of the director with arbitrary orientation 
$\mathbf{p}$ towards or away from the planar wall.  On the spherical surface 
defined by $|\mathbf{p}| = 1$, this corresponds to rotation of the director along a line 
of longitude, where $p_{z} = 1$ and $p_{z} = -1$ define the poles [Fig. 
\ref{fig:xi_rheotaxis_schematic}(c)]. 
The question, then, is whether, for a small perturbation $\Delta p_{x}$, activity will 
tend to drive the director towards the equator $p_{z} = 0$ of $|\mathbf{p}| = 1$ or 
towards the nearest pole. From Fig. \ref{fig:xi_rheotaxis_schematic}(c) we note that 
the effect of the small perturbation $\Delta p_{x}$ is to shift the tip of the director 
to a neighboring line of longitude. Therefore, if activity tends to rotate the director 
towards the equator, $\Delta p_{x}$ will increase (see 
Fig. \ref{fig:xi_rheotaxis_schematic}(c)) because the distance between neighboring 
lines of longitude increases as they approach the equator from the pole. The 
direction of this activity induced rotation (i.e., towards the pole or towards the 
equator) is determined by the value of $\xi^{*}$. For instance, consider a positive 
rheotactic state, for which $\pi < \xi^{*} < 3 \pi/2$.  In order to balance rotation 
by shear, activity must tend to rotate the director towards the pole $p_{z} = -1$, as 
shown in Fig. \ref{fig:xi_rheotaxis_schematic}(b) and (c). Hence, this dynamical fixed point 
is stable against small perturbations in $p_{x}$.  

\subsection{Calculation of the slip velocity}

Finally, we briefly outline the calculation (if necessary) of the slip velocity 
$\mathbf{v}_{s} = v_{s,\theta_p} \hat{\mathbf{e}}'_{\theta_p}$ (where 
$\hat{\mathbf{e}}'_{\theta_p}$ denotes the unit vector in the primed coordinate 
frame corresponding to the polar angle $\theta_{p}$ measured from the director, i.e., the symmetry axis of the particle) along the surface of the spherical particle which, for the specific swimmer models considered here, 
is an input to the problem outlined in Subsec. II.A.

For a ``squirmer'', $\mathbf{v}_{s}$ is specified by \textit{fiat} and does not 
depend on the configuration $(\mathbf{p},h)$ of the particle. Following Li and Ardekani 
\cite{li14}, we consider a squirmer for which the slip velocity is given by 
\begin{equation}
\label{eq:squirmer_slip}
v_{s,\theta_{p}} = B_{1} \sin(\theta_{p}) + B_{2} \sin(\theta_{p}) \cos(\theta_{p}),
\end{equation}
The model is characterized by the squirming mode amplitudes $B_{1}$ and $B_{2}$. In 
an unconfined quiescent fluid (i.e., \textit{f}ree \textit{s}pace) such a squirmer moves 
with velocity 
$U_{f.s.} = (2/3) B_{1}$.

A self-phoretic Janus particle generates solute molecules over the surface of a 
catalytic cap.  We parameterize the extent of the cap by $\chi_{0} = -\cos(\psi)$, where 
the angle $\psi$ is defined in Fig. \ref{fig:xi_rheotaxis_schematic}(a). It is assumed 
that the interaction between the active particle and the (non-uniformly 
distributed) solute molecules has a range much smaller than $R$ and $h$. Thus 
it can be accounted for as driving a tangential flow in a thin boundary layer of 
thickness $\delta << R, h$ surrounding the particle. This can be modeled as an effective 
slip velocity $\mathbf{v}_{s} = - b(\mathbf{r}) \nabla_{||} c(\mathbf{r})$, where 
$c(\mathbf{r})$ is the solute number density field \cite{anderson89,golestanian05}. The 
operator $\nabla_{||} = (\mathbf{1} - \mathbf{n} \mathbf{n}) \cdot \nabla$ is the 
projection of the gradient onto the active particle surface, with $\mathbf{n}$ 
denoting the surface normal oriented into the fluid.  The quantity $b(\mathbf{r})$ is 
the so-called ``surface mobility,'' determined by the details of the interaction between 
the active particle and the solute molecules. For instance, the sign of 
$b$ expresses whether the active particle is attracted to or repelled from the 
solute molecules. Throughout the present study, we assume repulsion. Since the 
solute number density $c(\mathbf{r})$ is affected by the presence of the wall, in this 
model the slip velocity $\mathbf{v}_{s}$ depends on the configuration $(\mathbf{p},h)$. 

If the effect of \textcolor{black}{advection} on $c(\mathbf{r})$ can be neglected, one can determine 
$c(\mathbf{r})$ (and with this $\mathbf{v}_{s}$) \textit{prior} to 
the consideration of the hydrodynamic problem posed in Subsec. II.A (see, e.g., 
Refs. \cite{anderson89, popescu09}).  (Within this approach it is disregarded 
how the flow 
field $\mathbf{u}$ couples back to $c(\mathbf{r})$.)  In this case, the solute number 
density is quasi-static, and obeys the Laplace equation $D \nabla^{2} c = 0$, subject to 
the boundary conditions for the normal derivatives $\hat{n} \cdot \nabla c = 0$ on the wall, $\hat{n} \cdot \nabla c = 0$ over the inert surface of the colloid, and $-D ( \hat{n} \cdot \nabla c ) = \kappa$ 
on the catalytic cap. The quantity $D$ is the diffusion coefficient of the solute 
molecules, and $\kappa$ is the flux of solute per unit area from the catalytically active cap. Concerning the first and second boundary conditions, the wall and the inert surface of the particle are taken to be completely impenetrable to solute molecules. In the third boundary condition, it is assumed that the rate of the reaction is independent of the local 
concentration of solute molecules.

It is valid to neglect \textcolor{black}{advective} effects if the P\'{e}clet numbers $Pe \equiv 
U_{f.s.} R/D$ and $Pe_{\dot{\gamma}} \equiv \dot{\gamma} R^{2}/D $ are small, where 
$U_{f.s.}$ is the particle velocity in \textit{f}ree \textit{s}pace. (The P\'{e}clet 
number $Pe$ is the ratio between the time scale for solute to diffuse a distance $R$ and 
the timescale for the particle to move a distance $R$. Similarly, the P\'{e}clet number 
$Pe_{\dot{\gamma}}$ characterizes the advection of the solute 
molecules by the external flow, for which the velocity scale is $\dot{\gamma} 
R$, relative to the diffusion of the solute molecules \cite{frankel14}.) 
Additionally, we assume that the Reynolds number $Re \equiv \rho U_{f.s.} R/\mu$ (where 
$\rho$ is the mass density of the suspending fluid and $\mu$ is its dynamic 
viscosity) is small, permitting the use of the Stokes equations to account for the 
hydrodynamics. For example, a self-propelled Janus colloid which catalyzes the 
decomposition of hydrogen peroxide into water and oxygen typically has a diameter of 
$\sim 5 \; \mathrm{\mu m}$ and moves with a velocity $\sim U_{f.s.} = 5 \; \mathrm{\mu 
m/s}$ through the aqueous solution ($\rho \simeq 10^{3}\; \mathrm{kg/m^{3}}$ and $\mu 
\simeq 10^{-3}\; \mathrm{Pa \, \times \, s}$). In this case, $Re \approx 10^{-5}$. Furthermore, 
estimating the diffusion coefficient of oxygen in water as $D \simeq 4 \times 
10^{-9} \; \mathrm{m^{2}/s}$, we obtain $Pe \approx 10^{-3}$ \cite{popescu10}. A 
representative shear rate in microfluidic devices is $\dot{\gamma} = 1\;\mathrm{s^{-1}}$ 
(see, for instance, Refs. \cite{kantsler14} and \cite{chengala13}), which 
renders $Pe_{\dot{\gamma}} \approx 10^{-3}$. Thus for typical active particles the 
above assumptions of small $Pe$, $Pe_{\dot{\gamma}}$, and $Re$ numbers are well justified.

\textcolor{black}{Before proceeding to an analysis of the numerical results, we remark on the physical realizations which would be captured by our model. In many experiments, the catalytic reactions involve more than one product as well as possibly 
a number of reactants. If the system remains in a reaction-rate limited regime (i.e., the reactants 
are in abundance and transported sufficiently fast so that there is no noticeable depletion near 
the catalyst), accounting for more than one product means to modify the expression for the 
slip velocity in a linear manner: the gradient of each reaction product ``i'' multiplied by its 
corresponding surface mobility $b_{i}$ is simply added in order to obtain the phoretic slip around the 
particle. In this case, the results presented here can be easily extended by 
replacing the effective number density $c$ by the sum over the densities of all products
with an effective $b$ such that $b \nabla_{||} c \rightarrow b_{1} \nabla_{||} c_{1} + b_{2} \nabla_{||} c_{2} + ...$. 
On the other hand, if the catalytic reaction is diffusion limited and involves at least two 
reactants, the source boundary condition at the catalytic-active region involves 
products of the densities of the reactants. In this case the equations describing the system 
become nonlinear, and the present model is neither applicable, nor does it lend itself to an obvious extension.}

\section{\label{numerics}Numerical Results and Discussion}

%%%%%%%%%%%%%%%%%%%%%%%%%%%%%%%%%%
\begin{figure}[htp]
\includegraphics{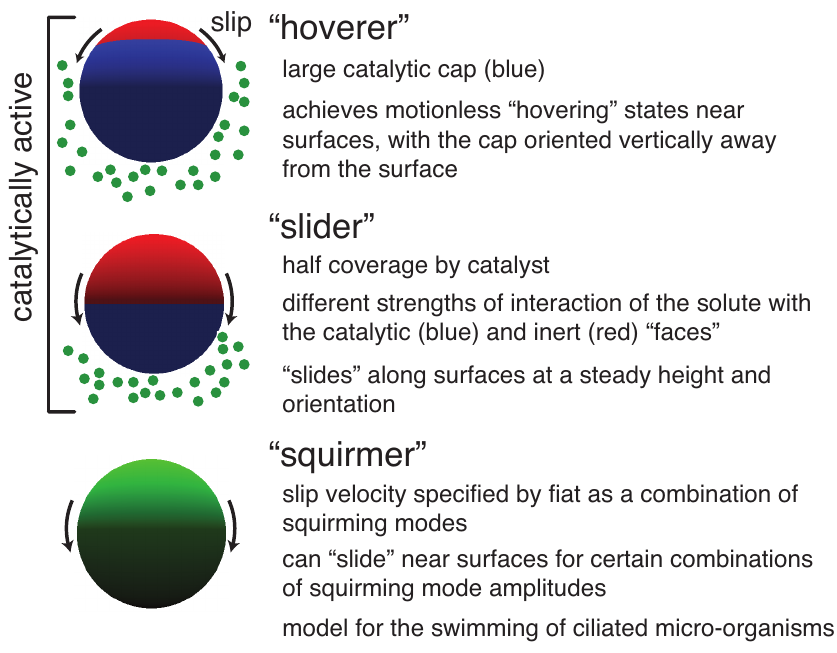}
\caption{\textcolor{black}{\label{particle_types} Illustration of the three types of spherical active particles studied here numerically. The three types differ in the kind of their surface activity. A catalytically active particle generates solute molecules (green discs) over the surface of a catalytic cap (blue). The interaction between the solute and the particle surface, here taken to be repulsive, drives a surface flow (slip velocity) as discussed in the main text, leading to a motion of the particle away from its cap. A ``hoverer'' (Sec. \ref{sect:hoverer}) exhibits a very high coverage by catalyst. As discussed in Ref. \cite{uspal14}, if it is near a planar wall it can achieve a motionless ``hovering'' state at a steady height above the wall and with the cap oriented away from the wall (as shown in Fig. \ref{hoverer_schematic}, left panel). For the hoverer considered here, we take the solute to interact identically with the catalytic and inert ``faces'' of the particle. In contrast, the ``slider'' considered in Sec. \ref{sect:slider} is only half covered by catalyst, but the solute interacts more strongly with its catalytic face (blue) than with its inert face (red). This particle tends to swim along surfaces at a steady height and orientation, or ``slides.'' Finally, for the ``squirmer''(Sec. \ref{sect:squirmer}), the slip velocity is not due to a self-generated  distribution of solute, but rather it is specified by fiat as a combination of ``squirming modes.'' For certain combinations of squirming mode amplitudes, the squirmer can achieve sliding state, too.}}
\end{figure}
%%%%%%%%%%%%%%%%%%%%%%%%%%%%%%%%%%

%%%%%%%%%%%%%%%%%%%%%%%%%%%%%%%%%%
\begin{figure}[htp]
\includegraphics{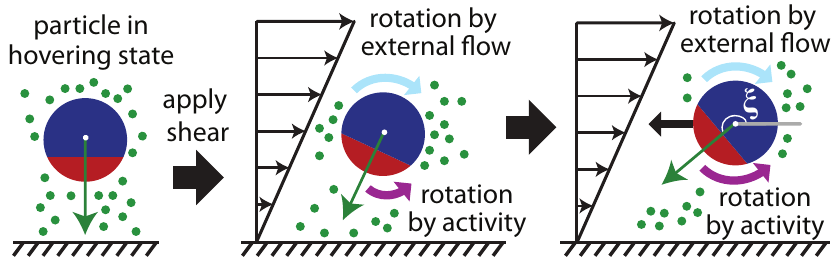}
\caption{\textcolor{black}{\label{hoverer_schematic}Schematic illustration of a dynamical process through which a ``hoverer'' particle can achieve a rheotactic state. Left panel: The particle is initially in a ``hovering'' state without any external flow. As discussed in Ref. \cite{uspal14}, the catalytic cap of the particle (blue) generates solute molecules (green discs), which interact with the particle surface via a repulsive potential. A ``cushion'' of repulsive solute in the space between the particle and the hard planar wall exactly balances the tendency of the particle to swim away from its cap. Center panel: An external shear flow tends to rotate the particle clockwise (cyan arrow). The near-wall chemical activity of the particle tends to rotate the particle back to the hovering orientation (magneta arrow). However, this contribution to particle rotation is small for a small displacement from the hovering orientation. The net effect is that the particle rotates away from the hovering orientation. Right panel: For sufficiently large displacements of the angle $\xi$ from the hovering state value $\xi = 3 \pi/2$, the two contributions to particle rotation balance, and a steady angle is achieved. Simultaneously, the particle still achieves a steady height above the wall due to the solute ``cushion'' effect. As a result, the particle translates upstream (black arrow), as its catalytic cap (blue) is oriented slightly downstream.}}
\end{figure}
%%%%%%%%%%%%%%%%%%%%%%%%%%%%%%%%%%

%%%%%%%%%%%%%%%%%%%%%%%%%%%%%%%%%%
\begin{figure*}[!htp]
\includegraphics[width=1.0\textwidth]{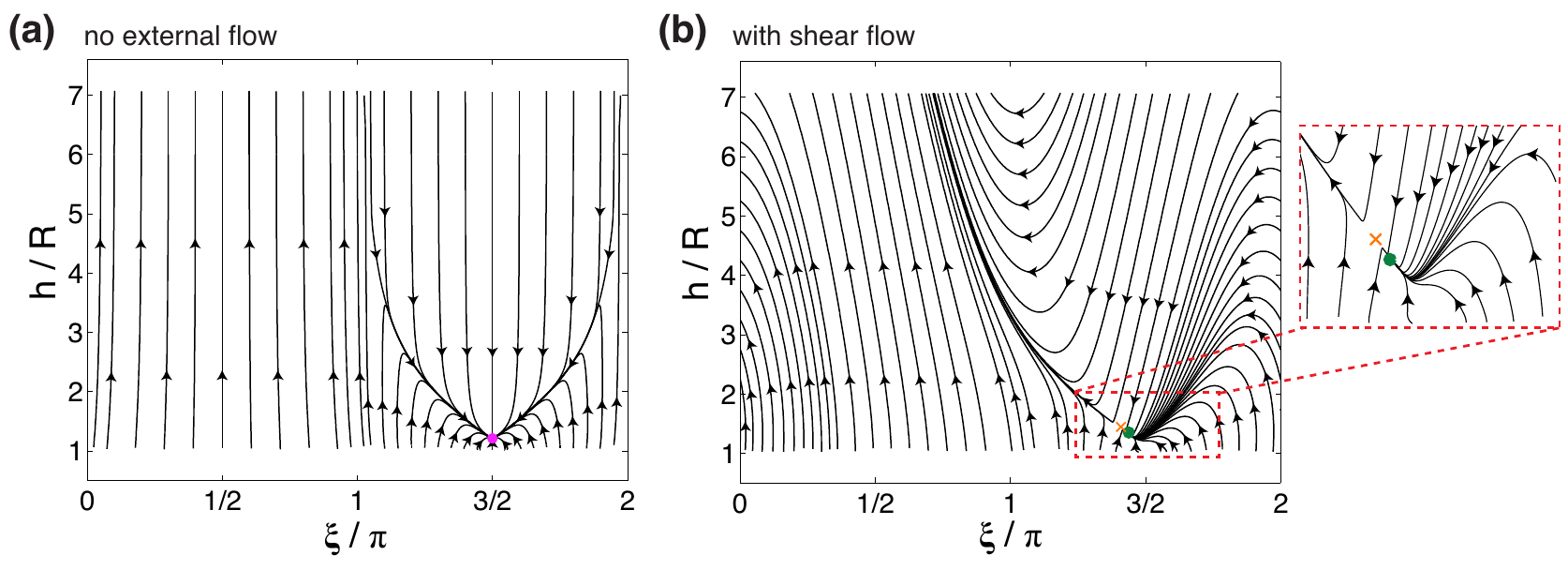}
\caption{
\label{fig:chi0_0.85_phase_plane} 
(a) Phase plane for a Janus particle with high coverage by catalyst ($\chi_{0} = 
0.85$) 
and uniform surface mobility if the director lies in the $yz$ plane ($p_{x} = 
0$) and if there is no external flow.  The filled magenta circle indicates an attractive 
``hovering'' 
state. In this state, the particle remains motionless at a height $h^{*}/R = 1.21$, with 
its cap 
oriented vertically and away from the wall ($\xi^{*} = 3 \pi/2 0 = 270^\circ$). The 
tendency of 
the particle to move away from its cap is balanced by the accumulation of solute near the 
wall.  
Clearly, the basin of attraction for this state is large. Additionally, many trajectories 
(roughly speaking, the whole basin of attraction) converge without overlapping to a 
``slow manifold'' (the two visible branches merging into the attractor), indicative of 
two 
timescale dynamics: the vertical motion is much faster than the rotation of the 
particle \cite{uspal14}.  Trajectories with an initial angle $0 < \xi_{0}/\pi < 
1$ escape from the wall. The mirror symmetry of the region $1 < \xi/\pi < 2$ across 
$\xi/\pi = 3/2$, as well as the mirror symmetry of the region $0 < \xi/\pi < 1$ 
across $\xi/\pi = 1/2$, is due to rotational symmetry around the $\hat{z}$ axis in the 
absence 
of flow.  
(b) Phase plane for the same Janus particle if the director lies in the shear 
plane 
($p_{x} = 0$) and in the presence of a shear flow with $\dot{\gamma} R/U_{0} = 0.02$. 
Note 
that 
the phase plane is periodic in the $\xi$ direction. Due to the rotation of the 
particle by 
the flow, the attractor (filled green circle) has migrated into the region 
$1 < \xi/\pi < 3/2$, and is now at $h^{*}/R = 1.35$ and $\xi^{*} = 257^{\circ}$.  As 
discussed in Sec. II, this is the region in which positive rheotaxis is possible.  
A particle 
in this state moves upstream with a steady height and orientation.  The orange cross indicates a saddle point. \textcolor{black}{An enlarged view of the region containing the saddle point and the attractor is provided in the right panel; some trajectories have been omitted for clarity.}
}
\end{figure*}
%%%%%%%%%%%%%%%%%%%%%%%%%%%%%%%%%%

When considering a Janus particle, we use the boundary element method (BEM) in order to solve the diffusion equation for the solute number density field $c(\mathbf{r})$. Additionally, for both propulsion mechanisms (the squirmer and the catalytically active particle), we use the 
BEM in order to solve the hydrodynamic subproblems I and II. A detailed introduction to the BEM is provided 
by Pozrikidis \cite{pozrikidis02}. We adopted his freely available \texttt{BEMLIB} 
code and modified it for the present study. For subproblem I, we 
calculate $f(h/R)$ and $g(h/R)$ for an (inactive) sphere in a shear flow at various 
heights $h/R$. We find that the numerically obtained values are in good agreement with the 
analytically obtained values given by Goldman \textit{et al} \cite{goldman67}. In a 
previous study, we provided a detailed description of how to apply the 
BEM in order to solve the diffusion equation (for a catalytic Janus particle) and 
subproblem II \cite{uspal14}. Within this approach we have obtained $U_{a,z'}$ and $\Omega_{a,x'}$ on a grid of $h$ and $\theta$.

In order to determine full particle trajectories for an initial condition 
$(h_{0}, \mathbf{p}_{0})$, we interpolate $f(h/R)$, $g(h/R)$, $U_{a,z'}$, and 
$\Omega_{a,x'}$ and integrate Eqs. (\ref{dot_px}) - (\ref{dot_h}) numerically. 
In the following, we restrict our numerical calculations to the region $1.02 \leq h/R 
\leq 7.1$.  Above $h/R = 7.1$, we consider the particle to have ``escaped'' the surface. 
Below $h/R = 1.02$, various of our physical approximations (e.g., that the effects of 
the solute distribution around the particle can be accounted for by a 
phoretic slip calculated within a thin layer, including that part of the surface of the 
particle which is in close proximity of the wall) are expected to break down.
For a catalytic Janus particle, $U_{0} \equiv |b| \kappa/D$ provides a characteristic 
velocity, and $T_{0} \equiv R/U_{0}$ a characteristic timescale. For instance, it 
was shown analytically that a catalytic Janus particle with uniform surface mobility and 
half coverage by catalyst has a velocity $U_{f.s.}/U_{0} = 1/4$ in \textit{f}ree 
\textit{s}pace \cite{popescu10}. When we take the catalytic cap and the inert region 
to have unequal surface mobilities, we use $U_{0} \equiv |b_{cap}| \kappa/D$. 
Similarly, for squirmers the amplitude $B_{1}$ of the first squirming mode provides a characteristic velocity scale, and the time $T_{s} \equiv R/B_{1}$ 
a characteristic timescale. The ratio between the squirming mode amplitudes 
defines a parameter $B_{2}/B_{1}$. In each of the cases which we shall study in this 
section, the corresponding velocity and time scales defined above and the particle radius $R$ will be employed to render the dynamical equations, Eqs. (\ref{dot_px}) - (\ref{dot_h}), dimensionless.

First, we shall apply the theoretical results derived in the previous section to show that the surface chemistry of a catalytically active Janus particle can be tailored such that it leads to the occurrence of positive (upstream) rheotaxis.  
We shall provide two rather distinct examples of such a design of the surface 
properties, each exploiting a particular pathway to produce the stabilizing wall-induced 
rotation component discussed in Subsec. \ref{section:linear_stability}. \textcolor{black}{These two examples are depicted schematically in Fig. \ref{particle_types}.}
In order to illustrate the generality of our theoretical results, we shall show 
that rheotaxis can occur for certain spherical squirmers.

Our approach to design is guided by the idea outlined in Fig. 
\ref{fig:xi_rheotaxis_schematic}(b). For positive rheotaxis, the particle director 
$\mathbf{p}$ must point upstream and towards the wall.  The external flow contributes a 
clockwise component, shown by the cyan arrow, to the angular velocity $\mathbf{\Omega}$ 
of the particle. For the particle to maintain a steady orientation, the near-wall 
swimming activity of the particle must contribute a counterclockwise component 
$\mathbf{\Omega}_{a} = -\mathbf{\Omega}_{f}$, shown by the magenta arrow, to 
$\mathbf{\Omega} = \mathbf{\Omega}_f + \mathbf{\Omega}_a$. Since the axisymmetric 
particle does not rotate in free space, the  counterclockwise component 
$\mathbf{\Omega}_{a}$ must be due to the effect of the wall on the fluid velocity field 
and the solute number density field. Additionally, the particle must be 
attracted to a steady height through its near-wall swimming activity. In other 
words, the tendency of the particle to swim away from its cap must be counteracted by the 
effect of the wall on the fluid velocity and solute density field.  In the following two 
subsections, we shall introduce two particle surface chemistries which fulfill 
these criteria.  By tailoring the surface chemistry, we can turn on or off, and 
rationally control, various physical mechanisms which contribute to the particle motion. 

\textcolor{black}{We note that the following three subsections (corresponding to the three particle designs shown in Fig. \ref{particle_types}) have a repetitive structure of arguments and presentation, so that the reader can choose to read the first subsection, skip ahead to the Conclusions, and return to read Subsections B and C at leisure.}

\subsection{Catalytic Janus particles with high coverage and uniform surface 
mobility \label{sect:hoverer}}

Previously \cite{uspal14} we studied the dynamics of a model catalytically active 
Janus particle suspended in a quiescent fluid and near a wall.  We found that, for 
a uniform surface mobility $b$, in the course of time a particle with a very high 
coverage by catalyst can be stably attracted to a ``hovering'' state in which it remains 
motionless at a height $h^{*}$ and angle $\xi^{*} = 3 \pi/2$ (see Fig. 
\ref{fig:xi_rheotaxis_schematic}(b)), i.e., with its catalytic cap oriented 
vertically and away from the wall \cite{uspal14}.  In this state, \textcolor{black}{depicted in the left panel of Fig. \ref{hoverer_schematic}}, the tendency of the particle to translate away from its cap (i.e., trying to avoid high solute concentrations) is 
balanced  by the accumulation of a solute ``cushion'' near the impenetrable wall 
(which is due to the confinement of the solute between the particle and the wall). 
The stability of this hovering state against perturbations in 
the $\hat{z}$ direction can be understood easily: if the particle moves closer to 
the wall, 
solute accumulation is enhanced, and repulsion from the wall is stronger; if the particle 
moves away from the wall, repulsion from the wall is weakened, but the particle still 
translates away from its cap.  Less obviously, hydrodynamic interactions with the wall 
stabilize the particle against perturbations in the angle $\xi$. Hydrodynamically, the 
particle is characterized as a ``puller'' (see the flow lines in Fig. 3(b) in Ref. 
\cite{uspal14}). Pullers are known to orient themselves perpendicular to planar surfaces 
via hydrodynamic interactions \cite{spagnolie12}.\footnote{Additionally, while our 
investigation of hoverers initially assumed uniform surface mobility, we have found that 
these mechanisms are preserved if the surface mobilities on the cap and 
on the inert regions are unequal, i.e., for a wide range of 
$b_{inert}/b_{cap} \neq 1$. Therefore the mechanism discussed here holds more 
generally.}

The phase plane for a hoverer with $\chi_{0} \, \equiv -\cos \psi = 0.85$ and uniform 
surface 
mobility $b$ in a quiescent fluid (i.e., no external flow) is shown in Fig. 
\ref{fig:chi0_0.85_phase_plane}(a). The phase plane indicates the evolution of the 
particle 
configuration $(h, \xi)$ for any initial condition $(h_{0}, \xi_{0})$. In the absence of 
flow, 
the system is symmetric for rotations around the $\hat{z}$ axis.  This rotational symmetry 
is 
evinced by two mirror symmetries in the phase plane: symmetry of the region 
$0 < \xi/\pi < 1$ for reflection across $\xi/\pi = 1/2$, and symmetry of the 
region $1 < \xi/\pi < 2$ for reflection across $\xi/\pi = 3/2$.  The ``hovering'' 
attractor is 
shown as a filled magenta circle at $\xi^{*} = 3 \pi/2$ and $h^{*}/R = 1.21$. As discussed 
in 
Ref. \cite{uspal14}, many trajectories in the basin of attraction are drawn to a 
``slow 
manifold'' indicative of two timescale dynamics: rotation is much slower than motion in 
$\hat{z}$ direction.  For initial conditions $\xi_0 < \pi$, the particle escapes 
the 
surface.  

%%%%%%%%%%%%%%%%%%%%%%%%%%%%%%%%%%
\begin{figure*}[!htp]
\includegraphics{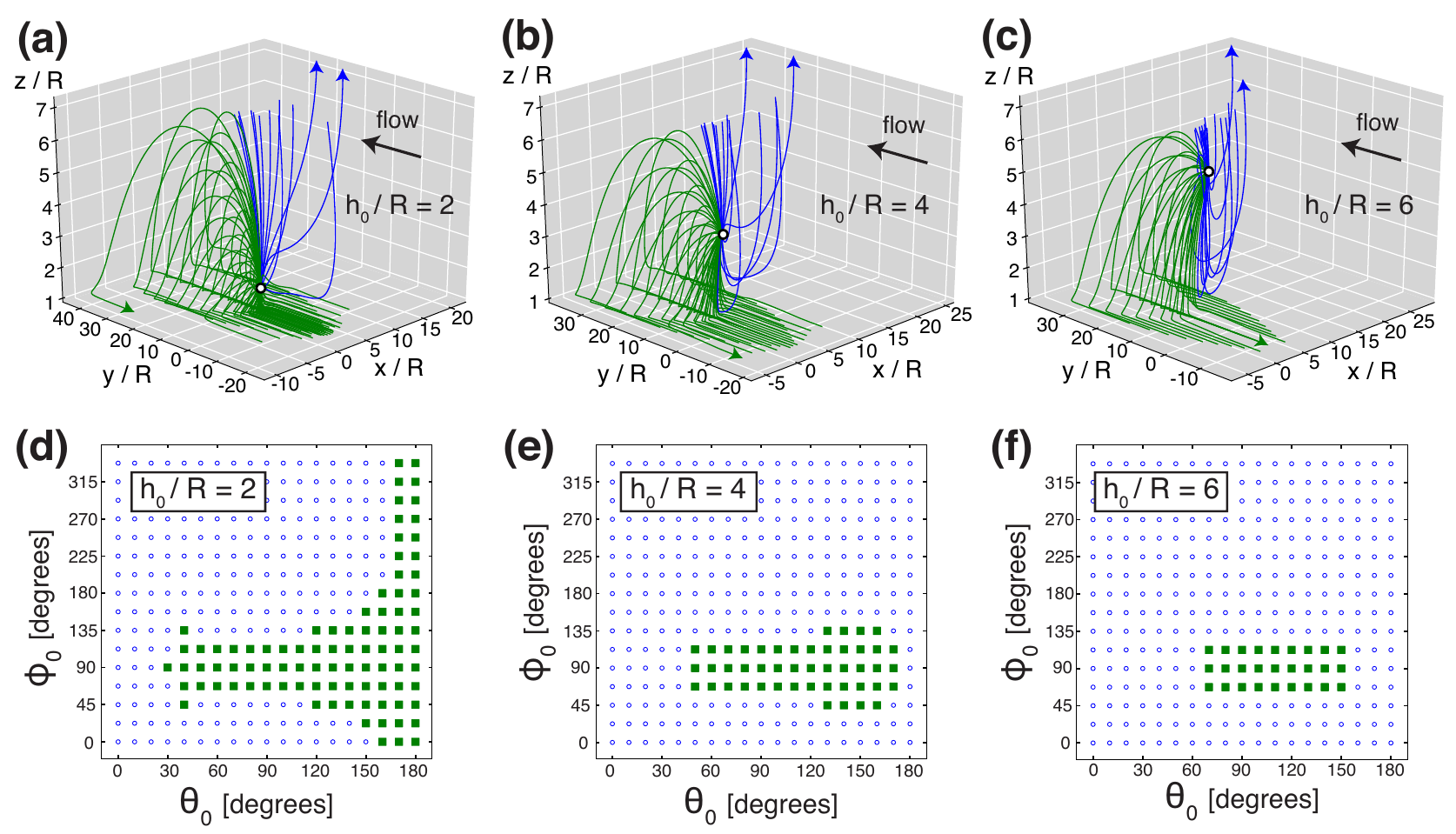}
\caption{
\label{fig:chi0_0.85_traj_phase_maps} 
Trajectories and phase maps for a catalytic Janus particle with high coverage by catalyst 
($\chi_{0} = 0.85$) and uniform surface mobility.  A trajectory starts from an 
initial 
height $h_{0}$, an initial director orientation $\theta_{0}$ and $\phi_{0}$, and 
an initial lateral position $(x_0,y_0) = (0,0)$. There 
is 
an ambient shear flow with dimensionless shear rate $\dot{\gamma} R/U_{0} = 0.02$. 
The panels (a), (b), and (c) show trajectories launched from $h_{0}/R = 2$, $h_{0}/R 
= 4$, 
and $h_{0}/R = 6$, respectively.  Green trajectories are attracted to a rheotactic 
swimming 
state, whereas blue trajectories escape from the surface. In each panel, a white circle 
shows the initial spatial position of all trajectories. Arrowheads on selected 
trajectories indicate the direction of motion. In panels (d), 
(e), and 
(f) we show phase maps which indicate the behavior for each initial height and 
orientation. The filled green squares correspond to the rheotactic trajectories, and 
the open blue circles correspond to the escaping trajectories. Note that, for visual 
clarity, not all of the trajectories considered in (d) through (f) are plotted in (a) 
through (c). In particular, escaping trajectories are only shown in the region $x > 0$.
}
\end{figure*}
%%%%%%%%%%%%%%%%%%%%%%%%%%%%%%%%%%

Now we consider the same ``hoverer'' in shear flow.  The flow will rotate the particle 
clockwise, 
decreasing $\xi$ from $\xi = 3 \pi/2$.  On the other hand, the hydrodynamic interaction 
which stabilized ``hovering'' will tend to rotate the particle back towards $\xi = 3 \pi/2$. \textcolor{black}{As depicted in the center and right panels of Fig. \ref{hoverer_schematic}, these two contributions to rotation can balance for sufficiently large angular displacement from the cap-down orientation. Hence the right panel of Fig. \ref{hoverer_schematic} provides a potential realization of the mechanism illustrated in Fig. \ref{fig:xi_rheotaxis_schematic}. Therefore we} 
anticipate that, at least for sufficiently slow external flows, there will a stable angle 
$\pi < \xi^{*} < 3 \pi/2$.  Moreover, we anticipate that the ``cushion'' effect will be preserved 
to 
produce a steady height $h^{*}$. The numerical simulations for, e.g., $\dot{\gamma} 
R/U_{0} = 
0.02$ lead to a phase plane as shown in Fig. \ref{fig:chi0_0.85_phase_plane}(b) and 
confirm these expectations: the attractor (filled green circle) is preserved and it 
migrates into the region $1 < \xi/\pi < 3/2$, being now located at $h^{*}/R = 
1.35$ 
and $\xi^{*} = 257^{\circ}$. Trajectories from a large section of phase space are drawn to 
this 
attractor. (Note that the phase plane is periodic in the $\xi$ direction.) Additionally, 
we 
indicate a saddle point (orange cross). The saddle point and attractor are 
rather close, as we have chosen a shear rate close to the (numerically estimated) 
upper critical value $\dot{\gamma}_{c} \approx 0.021 \times U_{0}/R$. At the critical 
value, 
the saddle point and the attractor collide and annihilate each other.  Above the 
critical 
shear rate, there is no stable rheotaxis, and, for this choice of the surface chemistry, 
all trajectories escape from the surface. We have chosen $\dot{\gamma} R/U_{0}$ near 
the 
upper critical shear rate since, as will be discussed in the Conclusions, ``strong'' 
external flows have the greatest experimental relevance and accessibility. Additionally, 
the relaxation time scale in Eq. (\ref{eq:px_exp_decay}) decreases as the shear rate is 
increased. Therefore, we expect the approach towards the rheotactic state to 
occur most rapidly for $\dot{\gamma} R/U_{0}$ being close to the upper 
critical shear rate. Although we leave an exhaustive parametric study to future 
research, we note that results of additional numerical calculations at lower 
values of $\dot{\gamma} R/U_{0}$, omitted here, agree with this expectation.

Since the attractor is in the region $\pi < \xi^{*} < 3 \pi/2$, it should be stable 
against perturbations of the director out of the shear plane, as discussed in Sec. II.  
In order to probe the stability of the attractor and its basin of attraction, we consider 
an ensemble of trajectories launched from various initial director angles $\theta_{0}$ and 
$\phi_{0}$ (see Fig. \ref{fig:schematic_3d}(a)), initial heights $h_{0}/R = 
2$, $h_{0}/R = 4$, and $h_{0}/R = 6$, and the initial lateral position $(x_{0}, y_{0}) 
= (0,0)$. For these three initial heights Figs. 
\ref{fig:chi0_0.85_traj_phase_maps}(a), (b), 
and 
(c) show three-dimensional trajectories of the coordinates $(x,y,z)$ of the centers of 
the 
particles. Each trajectory has been obtained from an initial position and 
orientation by numerically integrating Eqs. 
(\ref{dot_px})-(\ref{dot_h}). Green trajectories are ``rheotactic.'' Particles 
following these trajectories are attracted to the steady height 
$h^{*}/R = 1.35$ and angle $\xi^{*} = 257^{\circ}$ and move upstream.
Particles following blue trajectories escape from the surface 
(i.e., the trajectories cross $h/R = 7.1$ from below.) Phase maps, indicating how 
the particle behavior depends on the initial height and orientation, are shown in 
Figs. \ref{fig:chi0_0.85_traj_phase_maps}(d), (e), and (f), which correspond to 
$h_{0}/R = 2$, $h_{0}/R = 4$, and $h_{0}/R = 6$, respectively.  Clearly, rheotaxis is achieved for a large basin of initial conditions.

%%%%%%%%%%%%%%%%%%%%%%%%%%%%%%%%%%
\begin{figure*}[!htb]
\includegraphics[width=0.85\textwidth]{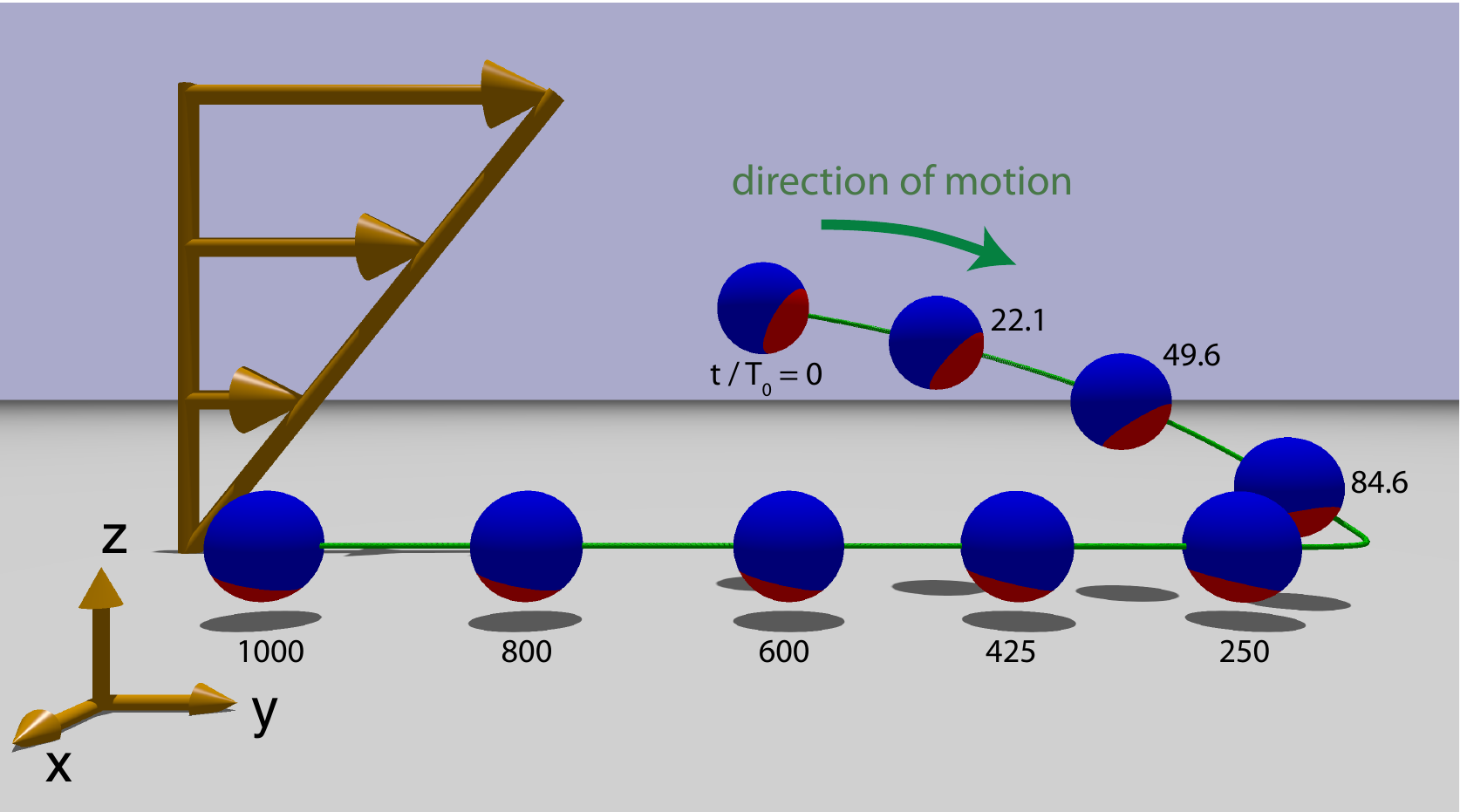}
\caption{
\label{fig:chi0_0.85_traj} 
Trajectory for a catalytic Janus particle with high coverage by catalyst 
($\chi_{0} = 0.85$, blue) and uniform surface mobility for the initial 
conditions 
$h_{0}/R = 6$, $\theta_{0} = 120^{\circ}$, and $\phi_{0} = 67.5^{\circ}$ in a shear flow 
with 
$\dot{\gamma} R/U_{0} = 0.02$. The particle is attracted to a configuration in which it is 
tilted 
slightly away from the ``hovering'' state (i.e., a completely vertical orientation) 
by the 
flow, and consequently moves upstream. Particle positions along the trajectory 
are 
labeled by the dimensionless times $t / T_{0}$ at which they occur. The particle is 
clearly almost in the rheotactic state at $t/T_{0} = 250$, which we estimate in the 
Conclusions to correspond to $t \sim 30 \, \textrm{s}$ for typical catalytic Janus 
particles used in experiments. Note that we have slightly enlarged the appearance of 
the inert part of the particle surface for visual clarity. For this 
trajectory, the time dependence of various quantities is shown by solid lines with stars 
in Fig. \ref{fig:chi0_0.85_rheo_plots}.
}
\end{figure*}
%%%%%%%%%%%%%%%%%%%%%%%%%%%%%%%%%%
%%%%%%%%%%%%%%%%%%%%%%%%%%%%%%%%%%
\begin{figure*}[!htb]
\includegraphics[width=0.85\textwidth]{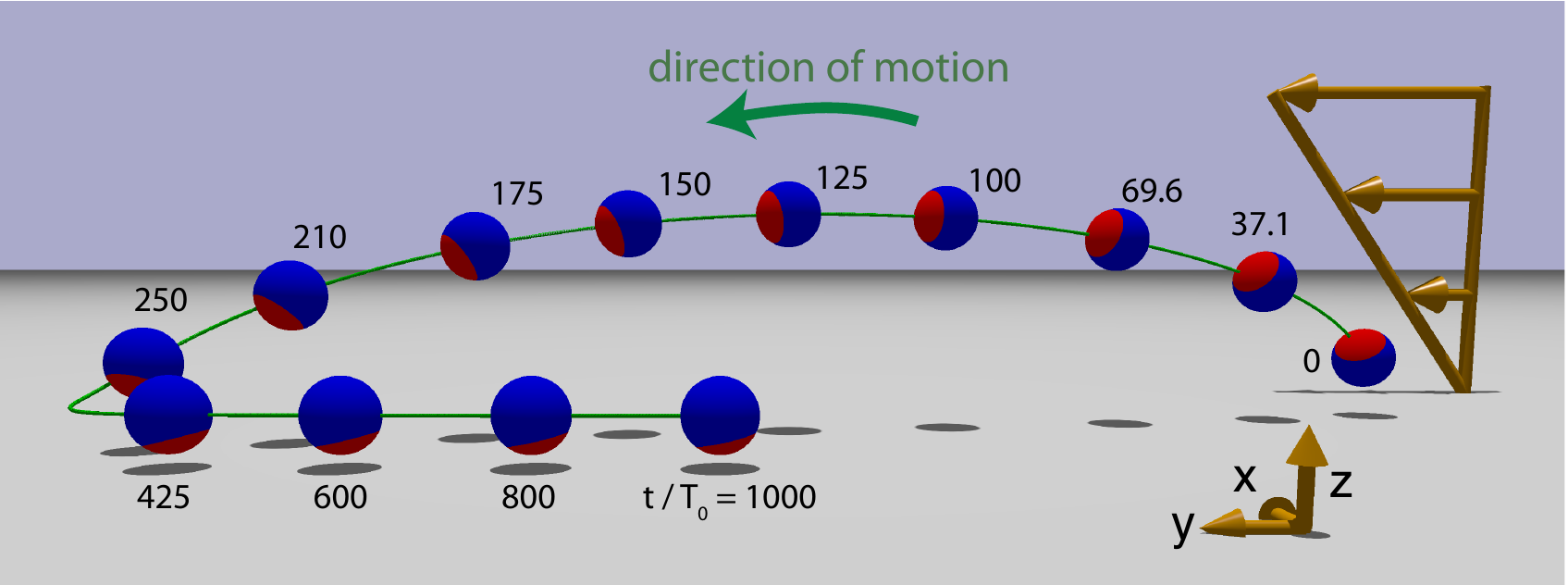}
\caption{
\label{fig:chi0_0.85_traj2} 
Trajectory for a catalytic Janus particle with high coverage by catalyst ($\chi_{0} = 
0.85$, blue) and uniform surface mobility for the initial conditions $h_{0}/R = 2$, 
$\theta_{0} = 40^{\circ}$, and $\phi_{0} = 135^{\circ}$ in a shear flow with 
$\dot{\gamma} R/U_{0} = 0.02$. The particle initially moves away from the wall, owing to 
the 
orientation of its catalytic cap.  A few particle diameters away from the wall, the 
influence of 
the wall is very weak. However, the particle is rotated by the flow to move back 
towards 
the wall.  As the particle approaches the wall, the influence of the wall strengthens.  
The 
particle rotates into the ``tilted hoverer'' configuration and moves upstream. Particle 
positions along the trajectory are labeled by the dimensionless times $t / 
T_{0}$ at which they occur. Note that we have slightly enlarged the appearance of the 
inert part of the particle surface for visual clarity. For this trajectory, 
the time dependence of various quantities is shown by solid lines with squares in Fig. 
\ref{fig:chi0_0.85_rheo_plots}.
}
\end{figure*}
%%%%%%%%%%%%%%%%%%%%%%%%%%%%%%%%%%

%%%%%%%%%%%%%%%%%%%%%%%%%%%%%%%%%%
\begin{figure*}[!htp]
\includegraphics[width=0.8\textwidth]{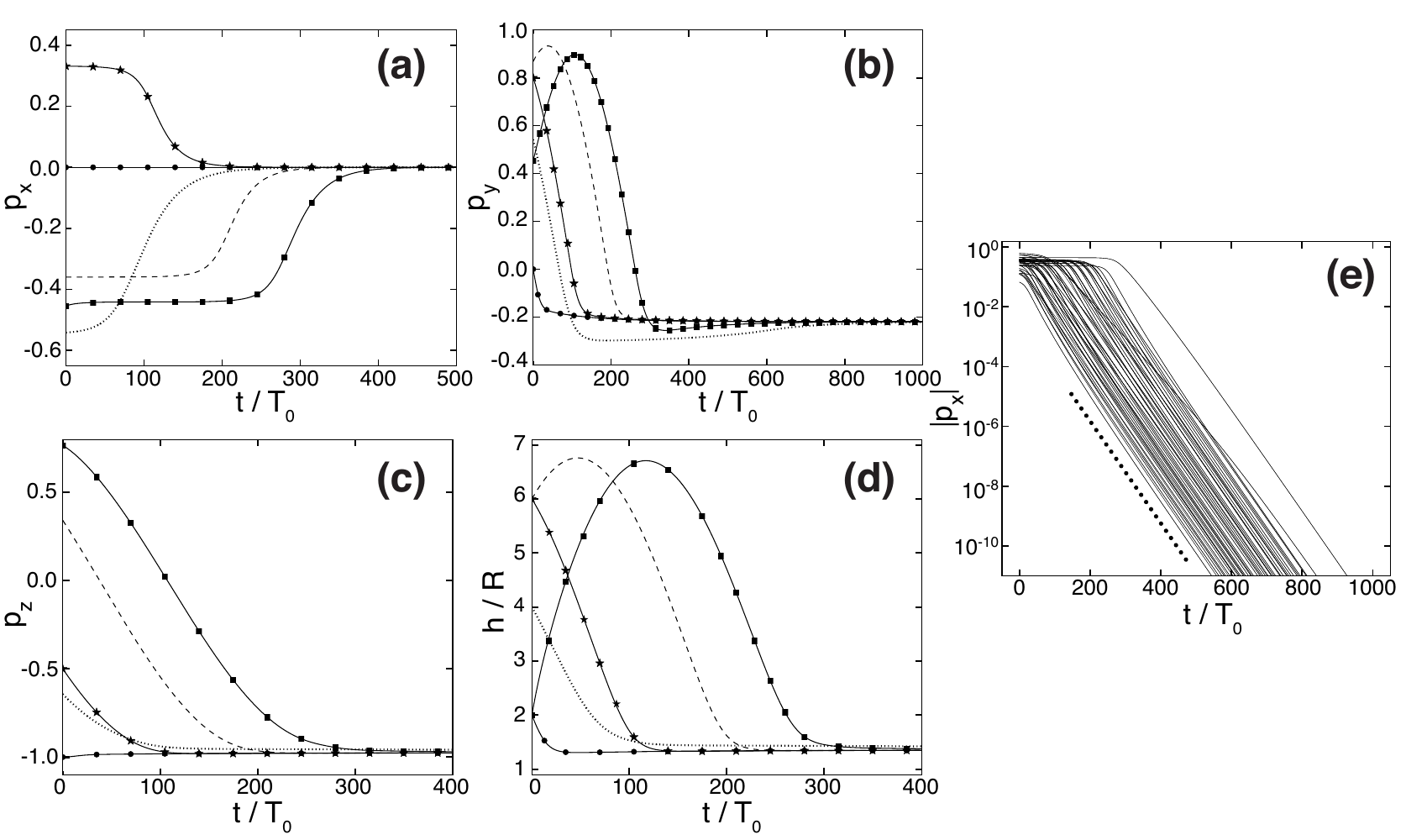}
\caption{
\label{fig:chi0_0.85_rheo_plots} 
(a)-(d) Time evolution of various quantities for five of the rheotactic 
trajectories of the catalytic 
Janus particle considered in Fig. \ref{fig:chi0_0.85_traj_phase_maps}. The quantity 
$p_{x}$ decays to 
zero, 
while $p_{y}$, $p_{z}$, and $h$ attain asymptotically non-zero values $p_{y}^{*} = 
-0.22$, 
$p_{z}^{*} = -0.98$, and $h^{*}/R = 1.35$. The initial conditions for these 
trajectories 
are: solid line with stars, $h_{0}/R = 6$, $\theta_{0} = 120^{\circ}$, $\phi_{0} = 
67.5^{\circ}$ (see also Fig. \ref{fig:chi0_0.85_traj}); 
solid line with filled circles, $h_{0}/R = 2$, $\theta_{0} = 180^{\circ}$, $\phi_{0} 
= 
0^{\circ}$; dotted line, $h_{0}/R = 4$, $\theta_{0} = 130^{\circ}$, $\phi_{0} = 
135^{\circ}$; 
dashed line, $h_{0}/R = 6$, $\theta_{0} = 70^{\circ}$, $\phi_{0} = 112.5^{\circ}$; solid 
line with 
squares, $h_{0}/R = 2$, $\theta_{0} = 40^{\circ}$, $\phi_{0} = 135^{\circ}$ (see also 
Fig. \ref{fig:chi0_0.85_traj2}). In panel (e), 
$|p_{x}|$ is plotted for all studied rheotactic trajectories on a 
semi-logarithmic scale. 
The exponential character of the decay of $\left| p_{x} \right|$ is clearly visible. 
The slope of the dotted black line indicates the exponential time dependence predicted 
by Eq. (\ref{eq:px_exp_decay}).
}
\end{figure*}
%%%%%%%%%%%%%%%%%%%%%%%%%%%%%%%%%%
An example of a rheotactic trajectory is shown in Fig. 
\ref{fig:chi0_0.85_traj}. 
Starting from the initial orientation $h_{0}/R = 6$, $\theta_{0} = 120^{\circ}$, and 
$\phi_{0} = 67.5^{\circ}$, the particle has nearly attained the steady height 
and 
orientation after moving only a few particle diameters.  The particle is attracted to 
a 
configuration in which it is tilted slightly away from the ``hovering'' state by the 
shear 
flow. 
As the (blue) cap is oriented slightly downstream (i.e., $\xi^* = 257^{\circ} < 
270^\circ$), 
the particle moves upstream.  In Fig. \ref{fig:chi0_0.85_traj2}, the particle 
starts near the wall, but pointing away from it.  Due to this initial orientation, the 
particle moves a few diameters away from the wall, where the hydrodynamic and chemical 
influence of 
the wall is very weak.  However, the flow rotates the particle to swim back towards the 
wall. Upon returning to the vicinity of the wall, the particle rotates into the ``tilted 
hoverer'' configuration.

For selected rheotactic trajectories, in Fig. \ref{fig:chi0_0.85_rheo_plots} we plot 
the 
time evolution of the particle height and of the components of the director. 
As expected, the component $p_{x}$ of the director perpendicular to the shear plane 
decays 
to zero. The height $h$ and the two in-plane components $p_{y}$ and $p_{z}$ attain 
asymptotically non-zero values. In Fig. \ref{fig:chi0_0.85_rheo_plots}(e), we plot 
the 
time evolution of $\left| p_{x} \right|$ for all rheotactic trajectories studied for 
Fig.  
\ref{fig:chi0_0.85_traj_phase_maps}. The decay of $\left| p_{x} \right|$  is clearly 
exponential, and the timescale for decay closely agrees with the prediction of Eq. 
(\ref{eq:px_exp_decay}), which is plotted as the dotted line in Fig. 
\ref{fig:chi0_0.85_rheo_plots}(e). Note that in 
Figs. \ref{fig:chi0_0.85_traj}, \ref{fig:chi0_0.85_traj2}, 
and \ref{fig:chi0_0.85_rheo_plots} time is given in dimensionless units as $t/T_{0}$. In 
a previous section, $T_{0}$ has been defined as $T_{0} \equiv U_0 /R$. 
As will be discussed in the Conclusions, we estimate $T_{0}$ to be  $T_{0} \approx 
0.125\,\textrm{s}$ for catalytic Janus particles as typically used in experiments.

\subsection{Catalytic Janus particle: half coverage and inhomogeneous surface 
mobilities\label{sect:slider}}

In order to demonstrate the general character of our theoretical findings, we seek 
to reach the rheotactic state of Fig. \ref{fig:xi_rheotaxis_schematic}(b) via the 
alternative pathway of a different surface chemistry designed to realize a distinct 
physical mechanism.  Specifically, we consider a particle which is half covered by 
catalyst ($\chi_{0} = 0$), but the surface mobilities of the inert region and of the catalytic cap are taken to differ: $b_{inert} \neq b_{cap}$.  In part, this choice is motivated by the fact that by default experimental studies use particles with half coverage, because these can readily be prepared by vapor deposition.  Moreover, since the catalytic and inert surface regions consist of different materials, they are likely to give rise to different surface mobilities, too.  
%%%%%%%%%%%%%%%%%%%%%%%%%%%%%%%%%%
\begin{figure*}[htp]
\includegraphics[width=1.0\textwidth]{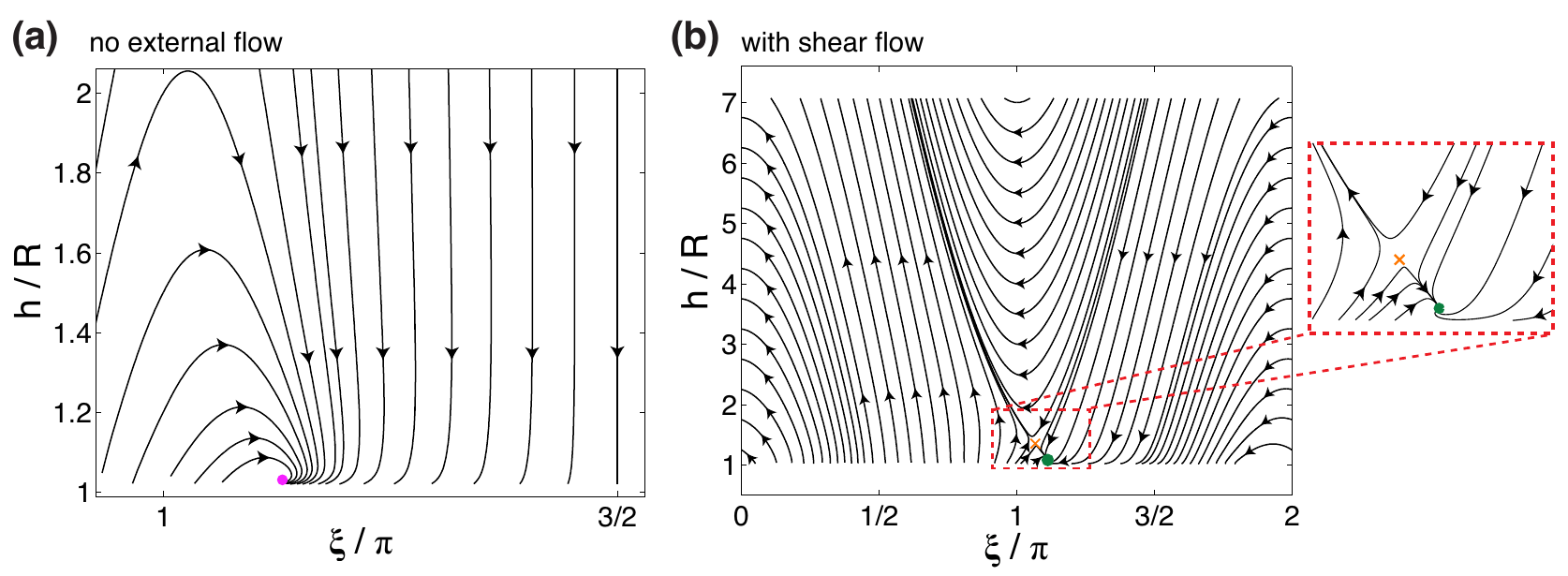} 
\caption{
\label{fig:inert_0.3_phase_plane} 
(a) Close-up of the phase plane of a Janus particle with half coverage ($\chi_{0} = 0$) 
and inhomogeneous surface mobilities $b_{inert}/b_{cap} = 0.3$ if the director is 
oriented in the $yz$ plane and there is no external flow. There is an attractor (filled 
magenta circle) at $h^{*}/R = 1.03$ and $\xi^{*} = 204^{\circ}$.  At this point, the 
hydrodynamic interaction with the wall, which tends to rotate the cap towards the wall, is 
balanced by the effect of wall-induced chemical gradients, which tend to rotate the cap 
away from the wall. Importantly for the behavior in the presence of flow (see (b)), 
there is a region in which the net effect of the activity of the particle is to rotate the 
cap away from the wall. Within the region $1 < \xi/\pi < 3/2$, this occurs where the phase 
space trajectories in (a) flow towards larger $\xi$ (i.e., to the right.)  (b) Phase plane 
for dynamics in the shear plane ($p_{x} = 0$) of the same particle with an external shear 
flow $\dot{\gamma} R/U_{0} = 0.05$.  There is a rheotactic attractor 
(filled green circle) at $h^{*}/R = 1.08$ and $\xi^{*} = 200^{\circ}$, as well as a 
saddle 
point (orange cross).  At the attractor, the net orientational effect of the chemical 
activity, which 
tends to rotate the cap away from the wall, is balanced by the orientational effect of 
the 
shear, which 
tends to rotate the cap towards the wall, as illustrated in Fig. 
\ref{fig:xi_rheotaxis_schematic}(b).  
Note that the phase plane is periodic in $\xi$. \textcolor{black}{An enlarged view of the region containing the saddle point and the attractor is provided in the right panel.}
}
\end{figure*}
%%%%%%%%%%%%%%%%%%%%%%%%%%%%%%%%%%

In our previous study of systems without flow, we have isolated two distinct 
wall-induced contributions to the rotation of a Janus particle \cite{uspal14}. One 
contribution is due to the hydrodynamic interaction of the particle with the wall. 
Disturbance flows created by the motion of the particle are reflected by the wall, 
coupling back to the particle. We found that, for half coverage ($\chi_{0} = 0$), 
hydrodynamic interactions always tend to rotate the catalytic cap 
of the particle towards the wall (and thus the director away from the wall). 
Therefore, this contribution to particle rotation cannot oppose the rotation by the 
shear flow (cyan arrow in Fig. \ref{fig:xi_rheotaxis_schematic}(b)) for $\pi < \xi < 3 
\pi/2$, i.e., it cannot provide the magenta arrow in 
Fig. \ref{fig:xi_rheotaxis_schematic}(b). However, we also found that if 
$b_{inert} \neq b_{cap}$, wall-induced chemical gradients can contribute to particle 
rotation. If $b_{inert}/b_{cap} < 1$, repulsion of the solute (i.e., the reaction product) 
from the inert region is weaker than repulsion from the catalytic cap. Accordingly, 
wall-induced chemical gradients (i.e., a higher concentration of solutes 
on the side of the particle surface closer to the wall) tend to rotate the catalytically 
active cap away from the wall. (Note that chemical gradients will not drive rotation of 
the particle in the absence of the wall because the Janus particle is axially 
symmetric.)  Therefore, taking $b_{inert}/b_{cap} < 1$ generates a contribution to the 
rotational velocity of the particle which corresponds to the magenta arrow in 
Fig. \ref{fig:xi_rheotaxis_schematic}(b).  

Therefore, we consider a particle with half coverage ($\chi_{0} = 0$) and 
$b_{inert}/b_{cap} = 
0.3$. A close-up of the phase plane for the dynamics is shown in Fig. 
\ref{fig:inert_0.3_phase_plane}(a) for the case that the director lies in the $yz$ 
plane and that there is no external flow. There is an attractor very close to the wall at 
$h^{*}/R = 1.03$ and $\xi^{*} = 204^{\circ}$. At this point, the hydrodynamic 
interaction with the wall, which tends to rotate the cap of the particle towards the wall, 
is balanced by the effect generated by wall-induced chemical gradients, which 
tends to rotate the cap away from the wall. The particle moves along the wall with a 
steady height and steady orientation, which we called a ``sliding'' steady state \cite{uspal14}.  Importantly, there are regions of the phase space in which the rotational effect of wall-induced chemical gradients is stronger than the rotational effect of the hydrodynamic interaction, such that in sum the cap tends to rotate away from the wall. In particular, in the interval $1 < \xi/\pi < 3/2$ in Fig. \ref{fig:inert_0.3_phase_plane}(a), the cap rotates away 
from the wall whenever trajectories flow towards larger $\xi$ (see Fig. 
\ref{fig:xi_rheotaxis_schematic}(b)), i.e., to the right of the plot in Fig. 
\ref{fig:inert_0.3_phase_plane}(a). In this region, rotation away from the 
wall by chemical activity is so strong that it can balance the rotational effect of shear 
towards the wall, as illustrated in Fig. \ref{fig:xi_rheotaxis_schematic}(b). In addition, 
$\dot{h} = 0$ whenever the tangent to the trajectories $h(\xi)$ in 
Fig. \ref{fig:inert_0.3_phase_plane}(a) is horizontal, i.e., $d[h(\xi(t))]/dt = 
\dot{\xi} \,(dh/d\xi) = 0$. 
Therefore, the swimming activity of the particle can potentially also on its 
own produce a steady height $h^{*}$.  We also note that some trajectories cross 
below the 
minimum height $h/R = 1.02$ which we consider. The particles seemingly ``crash'' 
into 
the wall.  It is likely that many of these trajectories recross the line $h/R = 1.02$ 
and 
flow to the attractor if the numerical calculations were extended below $h/R = 1.02$; 
however, as 
discussed above, the physical approximations inherent in the calculations break down 
very close to the wall.\footnote{\textcolor{black}{Without any 
major difficulties, the numerical calculations, within the current mathematical description, can be extended technically to separations below $h/R < 1.02$, corresponding to a particle/wall gap of less than $50\,\mathrm{nm}$ for a particle of radius $R = 2.5\,\mathrm{\mu m}$. However, these results would be physically questionable because our model cannot be expected to apply in that range. As discussed previously, for small particle/wall separations one can no longer assume a separation of length scales between solute/particle interactions and bulk hydrodynamic flow. Moreover, in that range, the typical surface interactions (such as van der Waals and electrostatic double layer forces) are relevant and therefore the assumption that the active particle is force and torque free breaks down. Such contributions can be included within a more complex model of active particles, which can be studied along similar lines.}}

%%%%%%%%%%%%%%%%%%%%%%%%%%%%%%%%%%
\begin{figure*}[htp]
\includegraphics{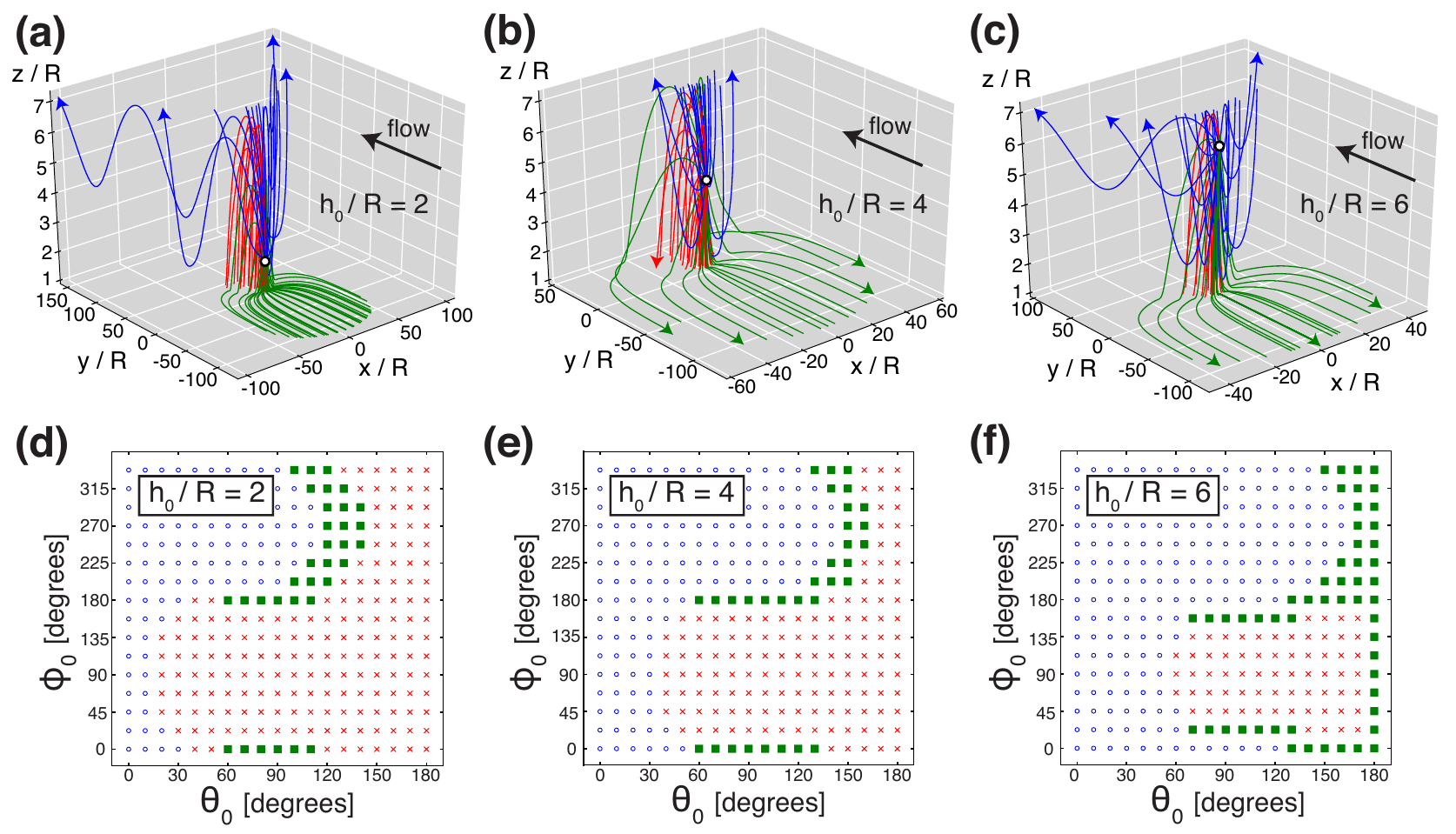}
\caption{
\label{fig:inert_0.3_traj_phase_maps} 
Trajectories and phase maps for a catalytic Janus particle with half coverage ($\chi_{0} 
= 
0$) and 
inhomogeneous surface mobilities $b_{inert}/b_{cap} = 0.3$. A trajectory starts 
from an initial height $h_{0}$, an initial director orientation $\theta_{0}$ and 
$\phi_{0}$, and the initial  
lateral position $(x_0, y_0) = (0, 0)$. There is an external shear flow with 
dimensionless shear 
rate $\dot{\gamma} R/U_{0} = 0.05$. The panels (a), (b), and (c) show trajectories 
launched from 
$h_{0}/R = 2$, $h_{0}/R = 4$, and $h_{0}/R = 6$, respectively.  Green trajectories are 
attracted to a rheotactic swimming state, 
blue 
trajectories escape from the surface, and red trajectories ``crash'' into the 
wall 
(i.e., they approach the surface closer than the minimal height we consider 
numerically). In each panel, a white circle shows the initial spatial position of all 
trajectories. Arrowheads on selected trajectories indicate the direction of motion. 
In panels (d), (e), and (f) we show phase maps which indicate the 
behavior for each initial height and orientation. The filled green squares correspond to 
the rheotactic trajectories; the open blue circles correspond to the escaping 
trajectories; and the red crosses correspond to the ``crashing'' trajectories.  Note that, 
for visual clarity, not all of the trajectories considered in (d) through (f) are plotted 
in (a) through (c).
}
\end{figure*}
%%%%%%%%%%%%%%%%%%%%%%%%%%%%%%%%%%
We now consider the same particle in shear flow.  The phase plane with shear rate 
$\dot{\gamma} 
R/U_{0} = 0.05$ is shown in Fig. \ref{fig:inert_0.3_phase_plane}(b). As anticipated, there 
is a 
rheotactic attractor (filled green circle), which is located at $h^{*}/R = 1.08$ and 
$\xi^{*} = 
200^{\circ}$.  Additionally, there is a saddle point (orange cross). As in the 
case of the ``hoverer,'' we have chosen a shear rate close to the upper critical 
value (numerically estimated to be $\dot{\gamma}_{c} R/U_{0} \approx 0.06$ for this surface 
chemistry), and therefore the saddle point and 
attractor are very close to each other.

%%%%%%%%%%%%%%%%%%%%%%%%%%%%%%%%%%
\begin{figure*}[htp]
\includegraphics[width=0.85\textwidth]{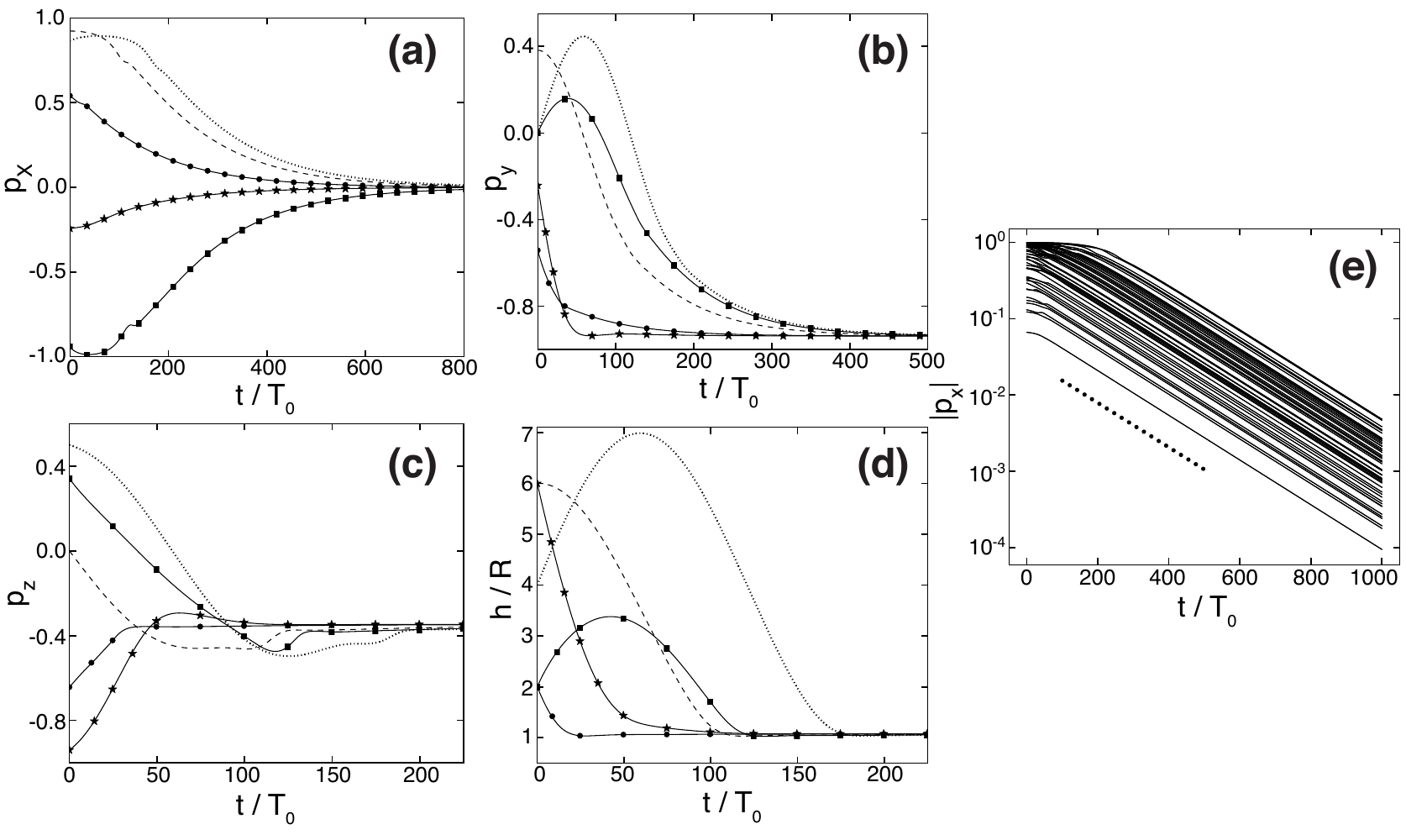}
\caption{
\label{fig:inert_0.3_rheo_plots} 
(a)-(d) Time evolution of $p_{x}$, $p_{y}$, $p_{z}$, and $h/R$ for five of 
the rheotactic 
trajectories considered in Fig. \ref{fig:inert_0.3_traj_phase_maps}. The quantity 
$p_{x}$ 
decays to 
zero, while 
$p_{y}$, $p_{z}$, and $h$ attain asymptotically non-zero values $p_{y}^{*} = 
-0.94$, 
$p_{z}^{*} = 
-0.35$, and $h^{*}/R = 1.08$.  The initial conditions for these trajectories are: 
solid line 
with stars, $h_{0}/R = 6$,  $\theta_{0} = 160^{\circ}$, $\phi_{0} = 225^{\circ}$; solid 
line with filled circles, $h_{0}/R = 2$, $\theta_{0} = 130^{\circ}$, $\phi_{0} = 315^{\circ}$; 
dotted line, 
$h_{0}/R = 4$, $\theta_{0} = 60^{\circ}$, $\phi_{0} = 0^{\circ}$; dashed line, $h_{0}/R = 
6$, 
$\theta_{0} = 90^{\circ}$, $\phi_{0} = 22.5^{\circ}$ (see also Fig. \ref{fig:inert_0.3_traj}); solid line with squares, $h_{0}/R = 
2$, 
$\theta_{0} = 70^{\circ}$, $\phi_{0} = 180^{\circ}$. In panel (e), $|p_{x}|$ is 
plotted for all studied rheotactic trajectories on a semi-logarithmic scale. The 
asymptotic decay of $\left| p_{x} \right|$ is clearly exponential. The slope of the 
dotted line indicates the exponential time dependence predicted by 
Eq. (\ref{eq:px_exp_decay}).
}
\end{figure*}
%%%%%%%%%%%%%%%%%%%%%%%%%%%%%%%%%%
%%%%%%%%%%%%%%%%%%%%%%%%%%%%%%%%%%
\begin{figure*}[htp]
\includegraphics[width=0.8\textwidth]{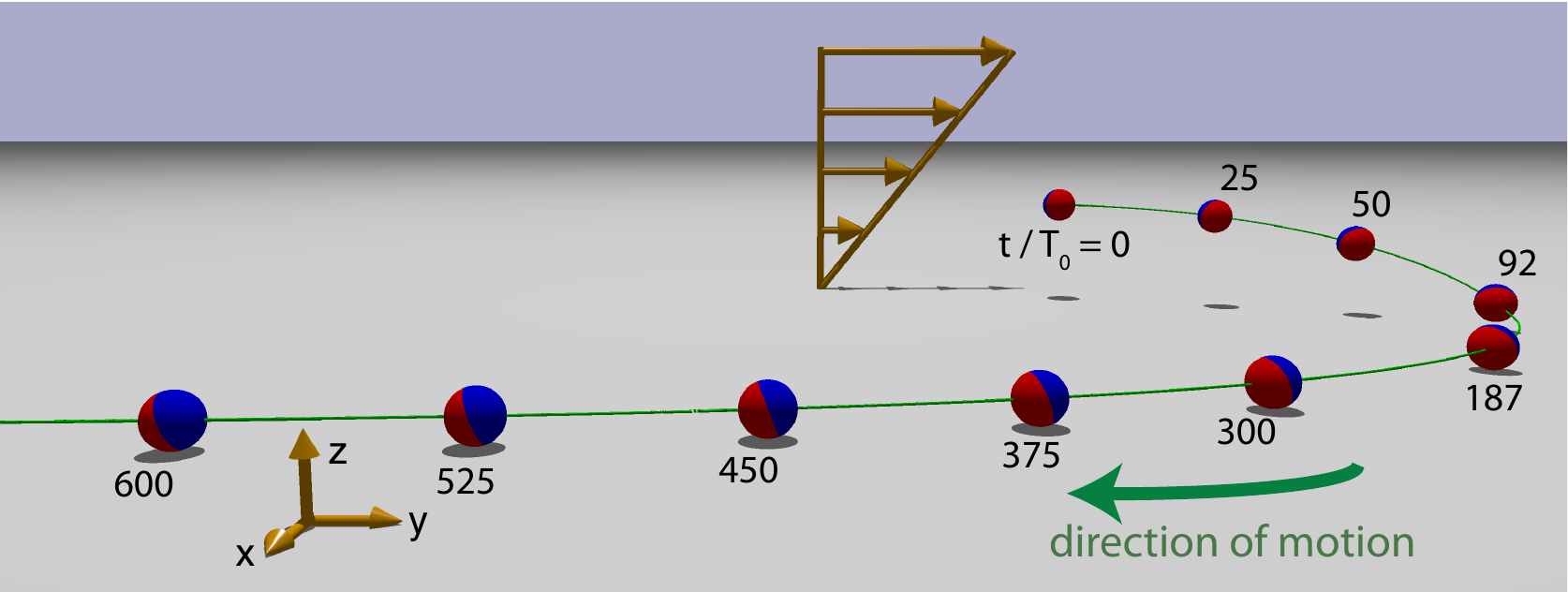}
\caption{
\label{fig:inert_0.3_traj} 
Trajectory for a catalytic Janus particle with half coverage ($\chi_{0} = 0$) and 
inhomogeneous surface mobilities $b_{inert}/b_{cap} = 0.3$ for the initial conditions $h_{0}/R = 6$, 
$\theta_{0} = 90^{\circ}$, and $\phi_{0} = 22.5^{\circ}$ in a shear flow with 
$\dot{\gamma} R/U_{0} = 0.05$. Particle positions along the trajectory are labeled by 
the 
dimensionless times $t / T_{0}$ at which they occur. The particle has ``turned 
around'' at $t/T_{0} = 350$, which we estimate in the Conclusions to correspond to 
$t \sim 45 \, \textrm{s}$ for typical catalytic Janus particles used in experiments. 
For this trajectory, the time dependence of various quantities is shown by dashed 
lines in Fig. \ref{fig:inert_0.3_rheo_plots}.
} 
\end{figure*}
%%%%%%%%%%%%%%%%%%%%%%%%%%%%%%%%%%

As for Fig. \ref{fig:chi0_0.85_traj_phase_maps}, we launch an ensemble of particles 
from the lateral position $(x_0,y_0)  = (0,0)$ of the center of the 
particles, for various initial
director 
orientations $\theta_{0}$ and $\phi_{0}$, and heights $h_{0}/R = 2$, $h_{0}/R = 4$, 
and 
$h_{0}/R = 6$. The three-dimensional trajectories of the center of the particles are 
shown in 
Figs. \ref{fig:inert_0.3_traj_phase_maps}(a), (b), and (c). The green trajectories 
are 
rheotactic. Particles following blue trajectories escape from the surface. Finally, the 
red 
trajectories are those which ``crash'' into the surface, as discussed above. Phase 
maps indicating the particle behavior as a function of the initial orientation are shown 
in 
Figs. \ref{fig:inert_0.3_traj_phase_maps}(d), (e), and (f). For selected rheotactic 
trajectories, in Fig. \ref{fig:inert_0.3_rheo_plots}(a)-(d) we plot the time evolution of 
the height $h$ and of the director components $p_{x}$, $p_{y}$, and $p_{z}$. In 
Fig. \ref{fig:inert_0.3_rheo_plots}(e), we plot $\left| p_{x} \right|$ for all rheotactic 
trajectories studied in Fig. \ref{fig:inert_0.3_traj_phase_maps}. As in Fig. 
\ref{fig:chi0_0.85_rheo_plots}(e), the decay time for $p_{x}$ closely agrees with the 
prediction of Eq. (\ref{eq:px_exp_decay}), plotted 
as the dotted line in Fig. \ref{fig:inert_0.3_rheo_plots}(e). Finally, a representative 
rheotactic 
trajectory is shown in detail in Fig. \ref{fig:inert_0.3_traj}.  The radius of curvature 
of the 
trajectory in its evolution towards the rheotactic steady state is clearly much larger 
than for the ``hoverer'' shown in Fig. \ref{fig:chi0_0.85_traj}. As with 
the hoverer, the dimensionless times $t/T_{0}$ in Figs. \ref{fig:inert_0.3_rheo_plots} and 
\ref{fig:inert_0.3_traj} can be converted to dimensional, and thus experimentally 
relevant values via the estimate $T_{0} \approx 0.125\,\textrm{s}$.

\subsection{Squirmer \label{sect:squirmer}}

In order to further reveal the general character of our theoretical results and 
predictions, we consider a spherical ``squirmer.'' A squirmer has a prescribed surface 
velocity.  
It interacts with a bounding wall hydrodynamically, but not chemically, because the 
surface 
flow is not driven by a distribution of solute.  The prescribed surface velocity is often 
taken to 
model the time-averaged motion of cilia on the surface of a micro-organism.

%%%%%%%%%%%%%%%%%%%%%%%%%%%%%%%%%%
\begin{figure*}[!htp]
\includegraphics[width=1.0\textwidth]{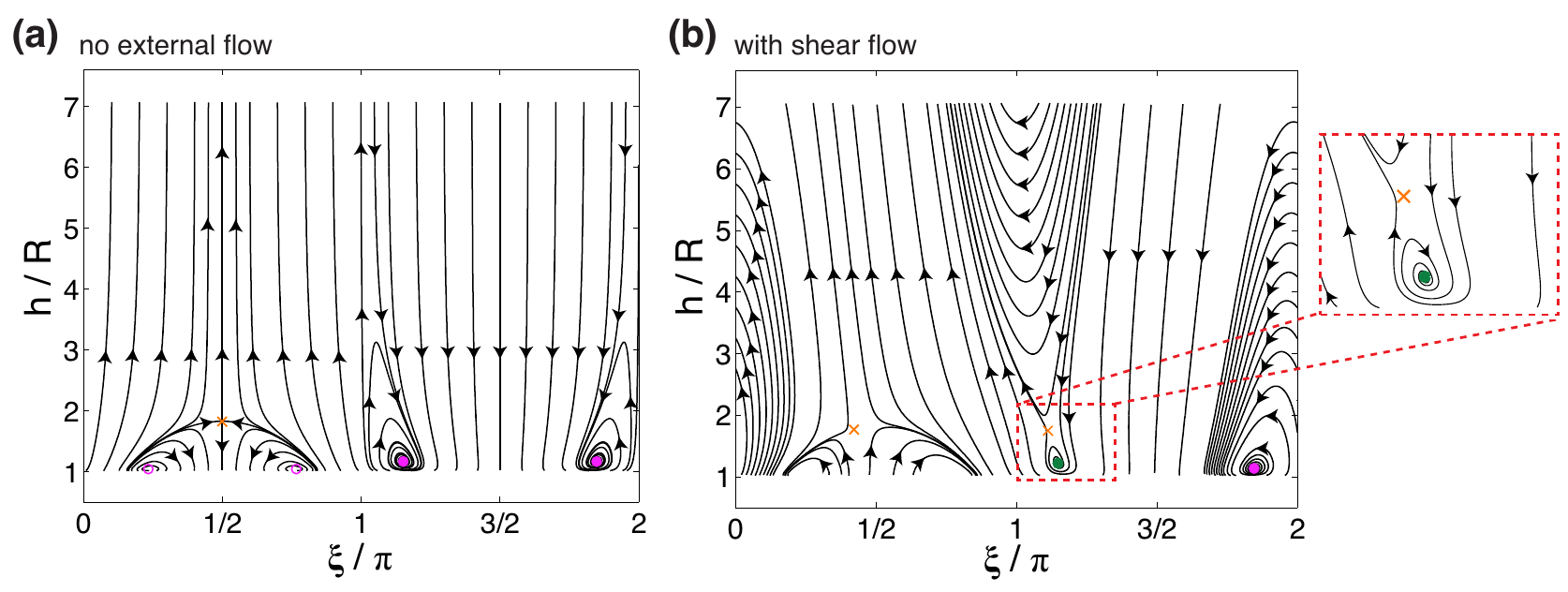}
\caption{
\label{fig:squirmer_phase_plane}
(a) Phase plane for a squirmer with $B_{2}/B_{1} = 7$ (Eq. (\ref{eq:squirmer_slip})) 
if the 
particle director lies in the $yz$ plane and if there is no external flow. There 
are 
stable dynamical attractors (filled magenta circles) for which the particle achieves 
a 
steady height and a steady orientation and ``slides'' along the wall. For these 
attractors, 
the particle director points towards the wall. There are also unstable fixed points 
(open 
magenta circles) and one saddle point (orange cross). Due to rotational symmetry around 
the 
$\hat{z}$ direction and in the absence of flow, the region $0 < \xi/\pi < 1$ is 
mirror 
symmetric across the line $\xi/\pi = 1/2$, and the region $1 < \xi/\pi < 2$ is 
mirror 
symmetric across the line $\xi/\pi = 3/2$. (b) Phase plane for the same squirmer if 
the particle director lies in the $yz$ plane and if there is a shear flow of 
strength $\dot{\gamma} R/B_{1} = 0.09$.  There is a rheotactic attractor (filled green 
circle). 
Since it lies in the region $1 < \xi/\pi < 3/2$, this attractor is stable 
against small 
perturbations of the director out of the $yz$ plane (i.e., for $p_{x} \neq 0$), as 
predicted in Sec. II.  There is another fixed point (filled magenta circle) which is 
stable 
against 
perturbations within the plane $p_x = 0$, but unstable against perturbations out 
of this plane. Likewise, this instability has been predicted in Sec. II for 
fixed 
points in the region $3/2 < \xi/\pi < 2$. There are two saddle points (orange 
crosses). The phase plane is periodic in the $\xi$ 
direction. \textcolor{black}{An enlarged view of the region containing the rheotactic attractor and the nearby saddle point is provided in the right panel.}
}
\end{figure*}
%%%%%%%%%%%%%%%%%%%%%%%%%%%%%%%%%%
We follow Li and Ardekani \cite{li14} in restricting our consideration to the first two 
squirming 
modes, given by 
$B_{1}$ and $B_{2}$ in Eq. (\ref{eq:squirmer_slip}). Li and Ardekani found that 
stable dynamical 
attractors exist for various values of $B_{2}/B_{1}$ and of the Reynolds number $Re$ 
\cite{li14}.  
(The lowest value of $Re$ they considered was $Re = 0.1$.) For instance, for 
$B_{2}/B_{1} = 3$ 
and $Re = 1$, they found an attractor with $h^{*}/R = 1.47$ and $\theta^{*} = 
103.2^{\circ}$. By construction our numerical method probes the limit $Re = 0$.  For 
$B_{2}/B_{1} = 3$ (and $Re = 0$), we have found an attractor at $h^{*}/R = 1.64$ and $\theta^{*} = 102.8^{\circ}$.  
We note that the dynamics of a squirmer near a boundary, including the possibility 
of moving at a 
stable height and orientation, was studied also by Ishimoto and Gaffney for 
$Re = 0$~\cite{ishimoto13}. 

Here, we consider a squirmer with $B_{2}/B_{1} = 7$.  We choose this large ratio in 
order
to achieve a strong hydrodynamic interaction with the wall, permitting rheotaxis for 
a 
relatively high shear rate $\dot{\gamma} R/B_{1} = 0.09$, as will be discussed below. 
 (By 
comparison, for $B_{2}/B_{1} = 3$ rheotactic states occur only for $\dot{\gamma} 
R/B_{1} < 
0.002$.) 
In Fig. \ref{fig:squirmer_phase_plane}(a) we show the phase plane for $B_{2}/B_{1} = 7$ 
if the 
director lies in the $yz$ plane and if there is no external flow. There are two 
stable 
dynamical attractors (filled magenta circles) for which the particle moves at 
a 
steady height 
and a steady orientation with its director pointing towards the wall, similar to the 
``sliding'' 
states we found for catalytic Janus particles \cite{uspal14}. Due to the rotational 
symmetry of subproblem II (see Fig. \ref{fig:chi0_0.85_phase_plane} and the corresponding 
discussion in 
Sec. III.A), these two attractors are actually the same in the sense 
that they 
correspond to the same sliding state, only that the particle slides towards the 
negative and positive $\hat{y}$ direction for the attractor with $\xi^*/\pi < 3/2$ and 
$\xi^*/\pi 
> 3/2$, respectively. In addition, there are two unstable fixed points with 
$\xi^*/\pi < 1$ (open magenta circles; this is again the same fixed point by symmetry) and a saddle point (orange cross).  

Now we consider the effect of an external flow with $\dot{\gamma} R/B_{1} = 0.09$ on the 
two 
attractors.  They remain attractors for \textit{in-plane} dynamics and stay at 
approximately the 
same locations in the phase plane, as can be seen in Fig. 
\ref{fig:squirmer_phase_plane}(b). 
The fixed point at $h^{*}/R = 1.22$ and $\xi^{*} = 207^{\circ}$  (filled green 
circle) lies in the region $1/2 < \xi/\pi < 3/2$ and therefore within the interval of 
$\xi$ for 
stable rheotaxis.  In this configuration, the director is oriented upstream and towards 
the wall, 
as in Fig. \ref{fig:xi_rheotaxis_schematic}(b). This fixed point is therefore a global 
attractor or 
an attractor for fully three-dimensional dynamics. For the other fixed point at $h^{*}/R = 
1.14$ and 
$\xi = 332^{\circ}$ (filled magenta circle), the particle director is oriented downstream 
($p_{y} > 0$) and towards the wall ($p_{z} < 0$); according to Eq. 
(\ref{eq:px_exp_decay}), this 
fixed point should be linearly unstable against perturbations out of the plane, i.e., 
away from $p_{x} = 0$. In addition, there are two saddle points (orange crosses).

%%%%%%%%%%%%%%%%%%%%%%%%%%%%%%%%%%
\begin{figure*}[htp]
\includegraphics{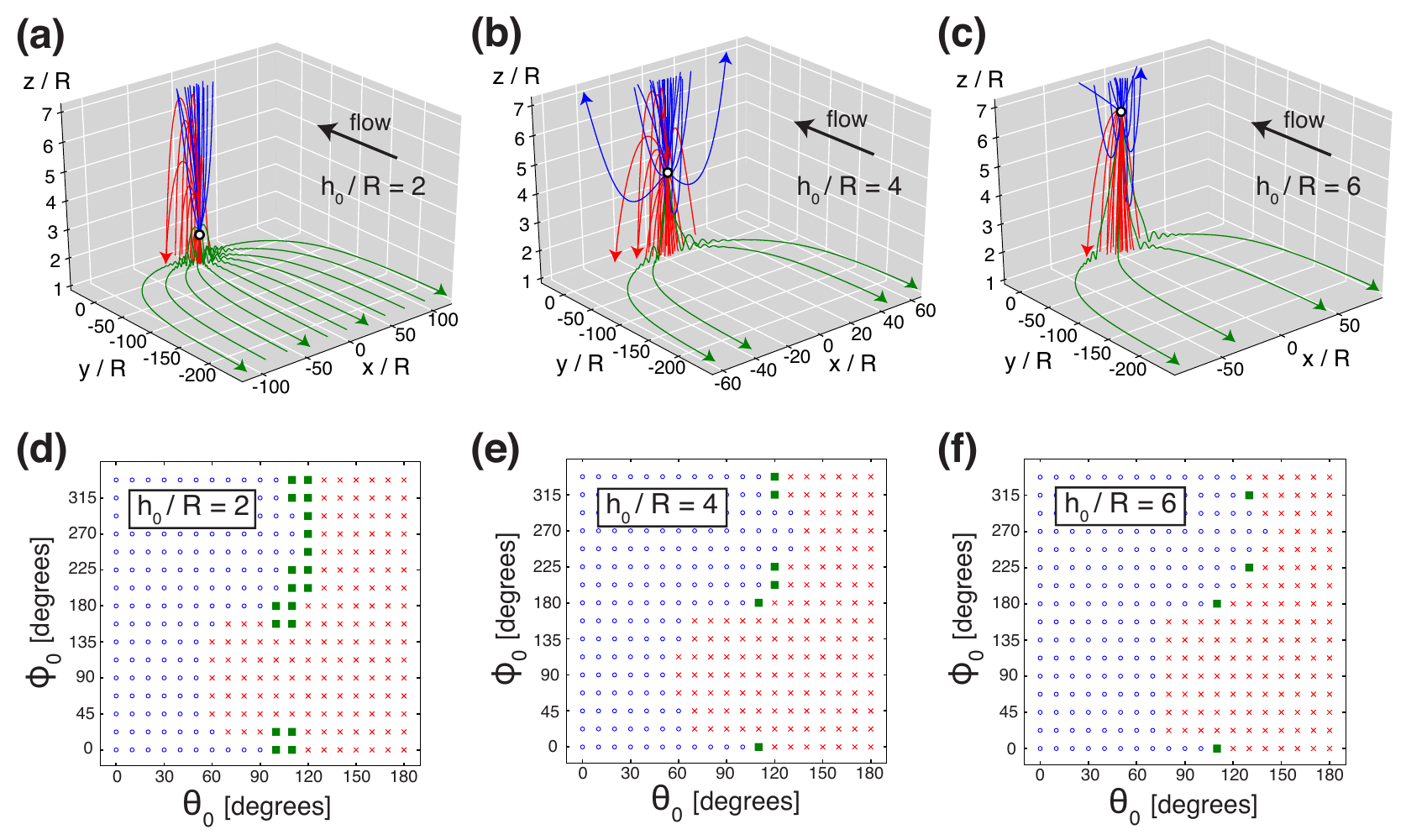}
\caption{
\label{fig:squirmer_traj_phase_maps} 
Trajectories and phase maps for a squirmer with $B_{2}/B_{1} = 7$ (Eq. 
(\ref{eq:squirmer_slip})).  
A trajectory starts from an initial height $h_{0}$, an initial director 
orientation 
$\theta_{0}$ and $\phi_{0}$, and the initial lateral position $(x_0,y_0) = (0, 
0)$. 
There is an external 
shear 
flow with dimensionless shear rate $\dot{\gamma} R/B_{1} = 0.09$. The panels (a), 
(b), and (c) 
show trajectories launched from $h_{0}/R = 2$, $h_{0}/R = 4$, and $h_{0}/R = 6$, 
respectively.  Green trajectories are attracted 
to a 
rheotactic swimming state; blue trajectories escape from the surface; and 
red trajectories ``crash'' into the wall (i.e., they approach the surface closer than 
the minimal height we consider numerically). In each panel, a white circle shows the initial spatial position of all trajectories. Arrowheads on selected trajectories indicate the direction of motion. In panels (d), (e), and (f) we show phase maps which 
indicate the behavior for each initial height and orientation. The filled green squares 
correspond to rheotactic trajectories; the open blue circles correspond to the escaping 
trajectories; and the red crosses correspond to the ``crashing'' trajectories. Note that, for 
visual 
clarity, not all of the trajectories considered in (d) through (f) are plotted in (a) 
through 
(c).
}
\end{figure*}
%%%%%%%%%%%%%%%%%%%%%%%%%%%%%%%%%

\begin{figure*}[!htp]
\includegraphics[width=0.5\textwidth]{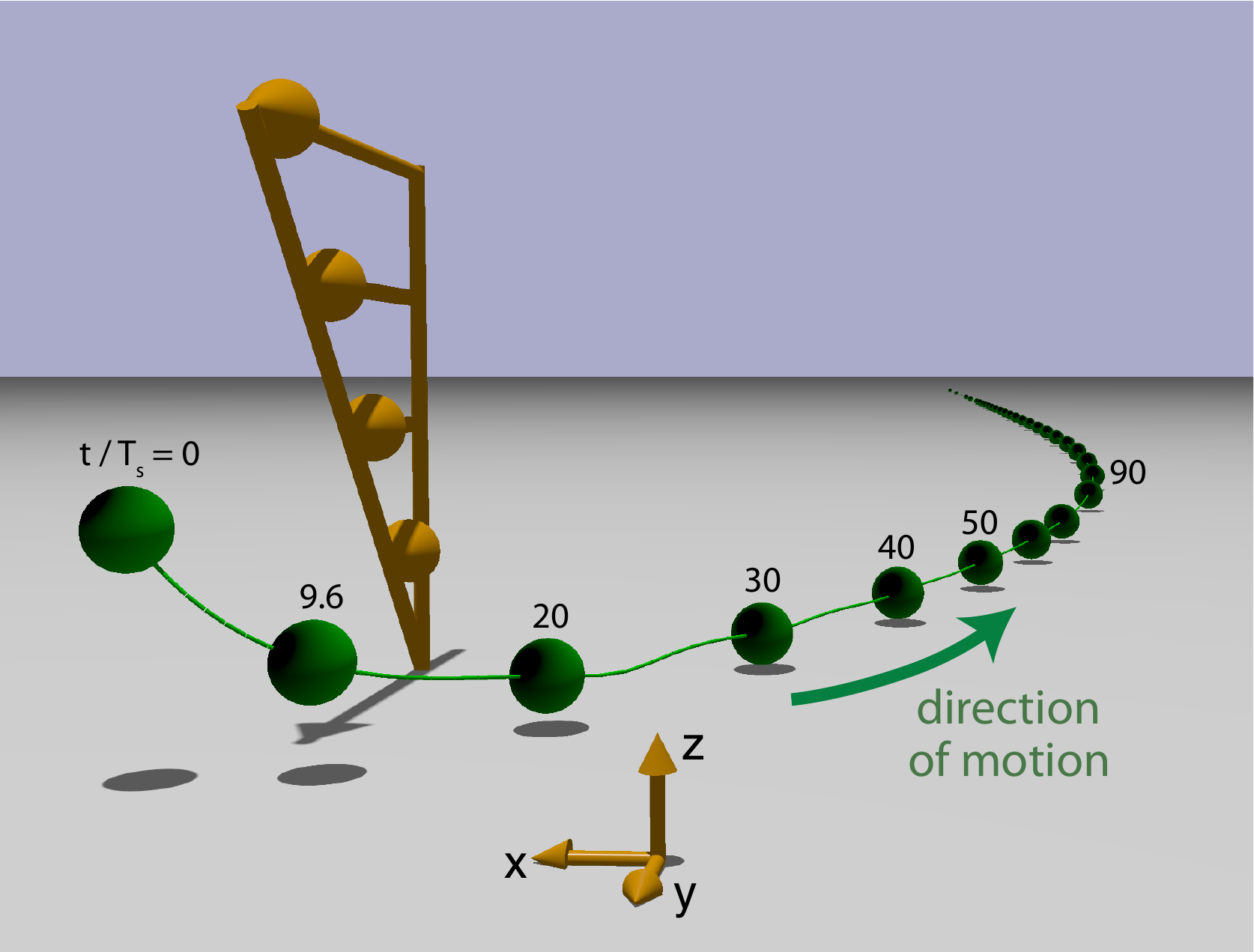}
\caption{
\label{fig:squirmer_traj} 
Rheotactic trajectory for a squirmer with $B_{2}/B_{1} = 7$ and initial condition 
$h_{0}/R = 6$, 
$\theta_{0} = 130^{\circ}$, and $\phi_{0} = 225^{\circ}$ in a shear flow with 
$\dot{\gamma} R/B_{1} 
= 0.09$. The black shaded part of the particle indicates its rear. The trajectory 
exhibits 
oscillations. Selected particle positions along the trajectory are labeled by the 
dimensionless times $t / T_{s}$ at which they occur. For this trajectory, the time 
dependence of various quantities is shown by solid lines with stars in Fig. 
\ref{fig:squirmer_rheo_plots}.
} 
\end{figure*}

As for Figs. \ref{fig:chi0_0.85_traj_phase_maps} and 
\ref{fig:inert_0.3_traj_phase_maps}, 
we launch an ensemble of particles from the lateral position $(x_0,y_0) = (0,0)$ 
of the center of the particles, for various initial angles $\theta_{0}$ and 
$\phi_{0}$, and heights $h_{0}/R = 
2$, 
$h_{0}/R = 4$, and $h_{0}/R = 6$. The  three-dimensional trajectories of these particles 
are shown 
in Figs. \ref{fig:squirmer_traj_phase_maps}(a), (b), and (c).  Green indicates 
rheotactic 
trajectories; blue indicates trajectories which escape from the surface; and 
red 
trajectories ``crash'' into the wall (i.e., they approach the wall closer than $h/R = 
1.02$, which is the minimum height we consider numerically). The phase maps in 
Figs. \ref{fig:squirmer_traj_phase_maps}(d), (e), and (f) indicate the particle 
behavior 
as a function of the initial height and the initial orientation. One 
rheotactic 
trajectory is shown in detail in Fig. \ref{fig:squirmer_traj}.
%%%%%%%%%%%%%%%%%%%%%%%%%%%%%%%%%%
%%%%%%%%%%%%%%%%%%%%%%%%%%%%%%%%%%

\begin{figure*}[!htp]
\includegraphics[width=0.8\textwidth]{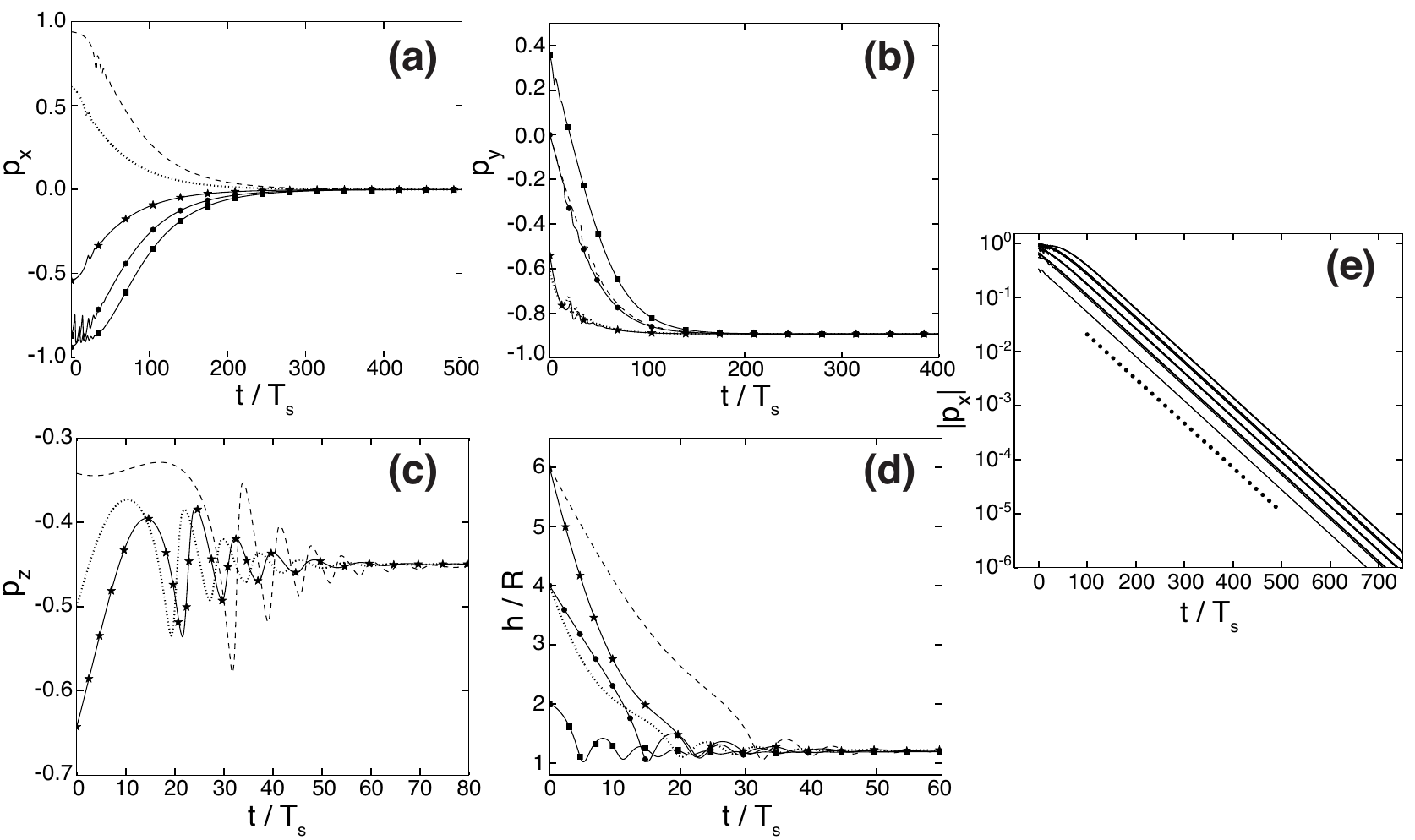}
\caption{
\label{fig:squirmer_rheo_plots} 
(a)-(d) Time evolution (in units of $T_s = R/B_1$) of various quantities for selected 
rheotactic 
trajectories of the squirmer discussed in Fig. \ref{fig:squirmer_traj_phase_maps}. 
The quantity 
$p_{x}$ decays to zero, while $p_{y}$, $p_{z}$, and $h$ attain asymptotically 
non-zero 
values 
$p_{y}^{*} = -0.89$, $p_{z}^{*} = -0.45$, and $h^{*}/R = 1.22$.   The initial conditions 
for the 
trajectories are: solid line with stars, $h_{0}/R = 6$, $\theta_{0} = 130^{\circ}$, 
$\phi_{0} = 
225^{\circ}$ (see also Fig. \ref{fig:squirmer_traj}); solid line with 
filled circles, $h_{0}/R = 4$, $\theta_{0} = 
110^{\circ}$, 
$\phi_{0} = 180^{\circ}$; dotted line, $h_{0}/R = 4$, $\theta_{0} = 120^{\circ}$, 
$\phi_{0} = 315^{\circ}$; dashed line, $h_{0}/R = 6$, $\theta_{0} = 110^{\circ}$, 
$\phi_{0} = 0^{\circ}$; solid line with squares, $h_{0}/R = 2$, $\theta_{0} = 
110^{\circ}$, 
$\phi_{0} = 157.5^{\circ}$. Note that, for visual clarity, in (c) only three 
trajectories are 
plotted. In (e),  $\left| p_{x} \right|$ is plotted on a semi-logarithmic scale 
for all rheotactic trajectories studied. The  slope of the dotted line indicates the 
exponential time dependence predicted by Eq. (\ref{eq:px_exp_decay}).
}
\end{figure*}

\begin{figure*}[htp]
\includegraphics[width=0.8\textwidth]{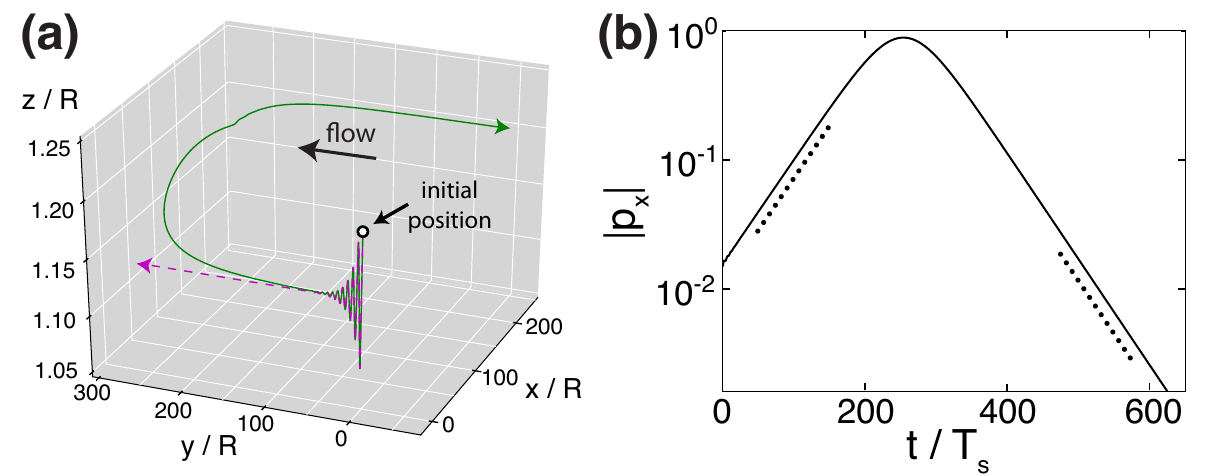}
\caption{
\label{fig:squirmer_unstable_point} 
(a) Two trajectories with initial height and initial orientation near the 
fixed point represented by the filled magenta circle in Fig. 
\ref{fig:squirmer_phase_plane}(b). 
This fixed point is a stable attractor for in-plane dynamics, i.e., for $p_{x} \equiv 0$, 
but it is unstable against small perturbations in $p_{x}$. In this configuration, the 
particle moves in the downstream direction with a steady height and orientation.  The two 
trajectories both have as initial conditions $h_{0}/R = 1.2$  and $\theta_{0} = 
150^{\circ}$, as well as the same initial lateral position $(x_0,y_0) = (0, 0)$ (white 
circle). However, one trajectory (magenta dashed line) has $\phi_{0} = 90^{\circ}$, i.e., 
initially $p_{x} = 0$, and therefore it is attracted to the downstream-moving state (i.e., 
the filled magenta circle in Fig. \ref{fig:squirmer_phase_plane}(b)). The other trajectory 
(solid green line) starts with $\phi_{0} = 89^{\circ}$, i.e., initially $p_{x} \neq 0$,  
and ultimately it is attracted to the rheotactic upstream-moving state (i.e., the filled 
green circle in Fig. \ref{fig:squirmer_phase_plane}(b)). (b) The time evolution of 
$|p_{x}|$ in units of $T_s = R/B_1$ for the rheotactic (
green) trajectory of (a), plotted on a semi-logarithmic scale. The slopes of the dotted 
lines indicate the exponential time dependences predicted by Eq. (\ref{eq:px_exp_decay}). 
}
\end{figure*}

For selected rheotactic trajectories, in Figs. \ref{fig:squirmer_rheo_plots}(a)-(d) 
we plot 
the time evolution of the height and of the orientation of the particle. Since 
the rheotactic 
attractor is 
oscillatory (see Fig. \ref{fig:squirmer_phase_plane}(b)), these quantities exhibit 
damped 
oscillations. The time evolution of $\left| p_{x} \right|$ for all rheotactic trajectories 
studied is plotted in Fig. \ref{fig:squirmer_rheo_plots}(e). As for the catalytically 
active
particles, the exponential decay of the component $p_{x}$ closely agrees with the one 
predicted by Eq. (\ref{eq:px_exp_decay}); in Fig. \ref{fig:squirmer_rheo_plots}(e) this theoretical prediction is indicated by the slope of the dotted black line. 
%%%%%%%%%%%%%%%%%%%%%%%%%%%%%%%%%%
%%%%%%%%%%%%%%%%%%%%%%%%%%%%%%%%%%
For the squirmer, we can test another prediction of the linear stability analysis.  As 
discussed previously in Fig. \ref{fig:squirmer_phase_plane}(b), there is a fixed point, represented by the filled magenta circle, which is, as obtained numerically, a stable 
attractor for in-plane dynamics, i.e., for $p_{x} \equiv 0$. At this fixed point, the 
particle moves \textit{downstream} at a steady height and a steady orientation. However, 
this fixed 
point lies in a region ($3/2 < \xi/\pi < 2$) for which linear stability analysis 
predicts 
fixed points to be unstable against small perturbations in $p_{x}$. In this 
context we launch two trajectories with initial conditions near this fixed point 
(Fig. \ref{fig:squirmer_unstable_point}). Both trajectories start with $h_{0}/R 
= 1.2$ 
and $\theta_{0} = 150^{\circ}$, but one with $\phi_{0} = 90^{\circ}$ (dashed 
magenta line in Fig. \ref{fig:squirmer_unstable_point}(a)), i.e., $p_{x} = 0$, and the other with $\phi_{0} = 89^{\circ}$ (solid green line), i.e., $p_{x} \neq 0$. For the trajectory 
with 
$\phi_{0} = 90^{\circ}$, the particle always remains in the shear plane ($p_{x} = 0$) and 
is 
attracted to the downstream-moving state. However, for the trajectory with $\phi_{0} 
= 
89^{\circ}$, the particle moves out of the shear plane, and is ultimately attracted 
to the 
rheotactic upstream-moving configuration. In Fig. 
\ref{fig:squirmer_unstable_point}(b), we 
show the time evolution of $\left| p_{x} \right|$ for the rheotactic trajectory.  Both the 
initial 
instability and the asymptotic approach to the rheotactic state clearly behave 
exponentially. 
The time dependences predicted by Eq. (\ref{eq:px_exp_decay}) are illustrated by dotted 
black lines. We note that the time scales for initial growth and asymptotic decay of 
$|p_x|$ are similar. This can be inferred from Fig. 
\ref{fig:squirmer_phase_plane}(b), in which the 
filled magenta circle and the filled green circle represent the downstream-moving and 
rheotactic upstream-moving fixed points, respectively. The original mirror symmetry 
across $\xi/\pi = 3/2$ of Fig. \ref{fig:squirmer_phase_plane}(a), in which there is no 
external flow, is approximately preserved, and hence the two fixed points have 
approximately the same values of $|p_{y}^{*}|$,  
$|p_{z}^{*}|$, and $h^{*}$. (As a reminder, Fig. \ref{fig:xi_rheotaxis_schematic}(a) 
relates $\mathbf{p}$ and $\xi$.) Since the time scale for growth or 
decay of $\Delta p_{x}$ is determined by $|p_{y}^{*}|$, $|p_{z}^{*}|$, and the local flow 
rate $\dot{\gamma} f(h^{*}/R)$ (see Eq. \ref{eq:px_exp_decay}), the two time scales are 
approximately the same for the two fixed points.
%%%%%%%%%%%%%%%%%%%%%%%%%%%%%%%%%%
%%%%%%%%%%%%%%%%%%%%%%%%%%%%%%%%%%

\section{Conclusions}

We have theoretically investigated the possibility that a spherical active particle 
with propulsion mechanism, which (i) is axisymmetric and (ii) can be described in terms 
of an effective slip velocity, may exhibit rheotaxis (and, in particular, upstream rheotaxis) in 
shear flow 
near a planar surface. (We define rheotaxis as to denote the approach of the particle 
to a 
robust and stable steady state in which the orientation vector lies within 
the plane of 
shear). We have found that rheotaxis of such a particle is indeed possible, even though its 
spherical 
geometry rules out the intuitive ``weather vane'' mechanism generally invoked in order 
to explain rheotaxis for \textit{elongated} microswimmers (e.g., sperm, \textit{E. coli}, 
and polymer/hematite dimers.)  

Furthermore, we have shown that for any such particles it is sufficient to analyze the 
dynamics of 
the particle with the orientation vector lying in the shear plane in order to 
determine 
whether or not rheotaxis occurs. In particular, for positive (upstream) rheotaxis, 
there must 
be an attractive steady state for the in-plane dynamics in which the orientation vector 
points upstream 
and towards the surface. For such a configuration, any small perturbation of the 
orientation vector 
\textit{out} of the shear plane is damped by the moving activity of the particle.  These 
findings 
significantly simplify the theoretical calculations. Moreover, they provide conceptually 
simple rules 
for the design of artificial rheotactic microswimmers. In order to achieve rheotaxis, 
one only needs to tailor the self-propulsion mechanism of the particle such that 
two criteria are satisfied: (i) the contribution of near-surface motion to the 
rotation of the particle can stably balance shear-induced rotation, as shown in 
Fig. \ref{fig:xi_rheotaxis_schematic}(b), and (ii) the particle, through its 
near-surface moving activity, can achieve a steady state of fixed height.  Specific 
examples 
of design for rheotaxis have been provided by exploring two distinct pathways for 
fulfilling these criteria 
via tailoring the surface chemistry of a catalytically active Janus particle. In the 
first pathway, the phoretic mobility is distributed homogeneously over the surface of the 
active particle, and the coverage by catalyst is adjusted to provide the required 
activity-induced rotation. In the second pathway, 
the 
coverage by catalyst is kept to the value of one half, which is simpler for experimental 
realizations, but the phoretic mobility is taken to be distributed inhomogeneously over 
the 
surface of the active 
particle.
In both these cases, the numerical solutions of the equations of motion evidence 
rheotaxis and 
confirm the analytical prediction for the decay time for the component of the 
orientation 
vector out of the shear plane.

The numerical study we presented in Sec. \ref{numerics}.A and B revealed the existence 
of a threshold value $r_c$ of the dimensionless parameter $r = \dot{\gamma} R/U_0$ above which 
rheotaxis 
is no longer possible. It is therefore important to understand the consequences of this 
constraint 
in the context of experimental studies with catalytically active Janus particles. We can 
judge the feasibility of experimentally realizing rheotactic Janus active particles by 
estimating the values of the shear rates corresponding to the threshold values $r_c = 0.021$ 
and 
$r_c = 0.06$ for ``hoverer'' (Sec. \ref{numerics}.A) and ``slider'' (Sec. 
\ref{numerics}.B), respectively, which have been determined by our numerical 
calculations. 
Considering, as in Sec. \ref{theory}.E, a typical Janus particle of radius $R \approx 
2.5\; \mathrm{\mu m}$ which is half-covered with catalyst and in the free space moves with 
a velocity $U_{f.s.} \approx 5\; 
\mathrm{\mu 
m/s}$, i.e., $U_{0} = 4 \times U_{f.s.} \approx 20 \; \mathrm{\mu m/s}$ \cite{popescu10}, 
we find that the maximal shear rate, above which rheotaxis is no longer possible, is 
$\dot{\gamma}_c = r_c U_0/R = 0.17 \; 
\mathrm{s^{-1}}$ for a ``hoverer'' and $\dot{\gamma}_c = 0.48 \; \mathrm{s^{-1}}$ for a 
``slider'', respectively. These values are low, but within or nearly within the range 
$\dot{\gamma} = 0.2 \; \mathrm{s^{-1}}$ to $\dot{\gamma} = 9 \; \mathrm{s^{-1}}$ explored 
in Ref. 
\cite{kantsler14}, and within the range $\dot{\gamma} = 0.1 \; \mathrm{s^{-1}}$ to 
$\dot{\gamma} = 
20 \; \mathrm{s^{-1}}$ explored in Ref. \cite{chengala13}. In order to relax these 
bounds on 
the shear rate, the ratio $R/U_0$ must be increased. We suggest two possible approaches to 
achieve 
this. The first one is to exploit the nonlinear relationship between size and propulsion 
velocity in the range of small particle 
radii, 
which was reported by Ebbens \textit{et al.} \cite{ebbens12}. Following Ref. 
\cite{ebbens12}, a 
$1 \; \mathrm{\mu m}$ radius particle with half coverage has a moving velocity of 
$U_{f.s.} = 9\;\mathrm{\mu m/s}$ \cite{ebbens12}.  For such an active Janus 
particle, half-covered by catalyst, the maximal shear rate, which corresponds to a slider 
and allows for rheotaxis, becomes $\dot{\gamma}_c \simeq 2.2 \; \mathrm{s^{-1}}$. However, 
this estimate should be considered with due care. As the size of the active particle 
decreases, the effects of the thermal noise on 
the 
orientation of the particle, which have been neglected in the present study, become 
significant and therefore the assumptions of the model leading to the predictions of the 
values for $r_c$ will break down. A second approach would be to exploit the fact 
that the velocity scale $U_0$ increases with the fuel concentration (at least within a 
certain 
range); this suggests using a high fuel concentration. For instance, Baraban \textit{et 
al.} report that a $R = 2.5 \; 
\mathrm{\mu m}$ active Janus particle with half coverage can achieve speeds $U_{f.s.}$ of 
more 
than $8 \; \mathrm{\mu m/s}$ at a high concentration ($15\%\;$ volume fraction) of 
hydrogen 
peroxide \cite{baraban12}. In this case, the maximal shear rate, which corresponds to a 
slider and at which rheotaxis can occur, would be in the range of 
$0.8 \; \mathrm{s^{-1}}$.

Having estimated dimensional values of $U_{0}$ and $R$ for a catalytic particle, we 
obtain 
the time scale $T_{0} = R/U_{0} = 0.125 \; \mathrm{s}$.  We can acquire a rough sense of 
the time required for rheotaxis in an experiment by converting the dimensionless times 
$t/T_{0}$ given by the black numbers in Figs. \ref{fig:chi0_0.85_traj},  
\ref{fig:chi0_0.85_traj2}, and \ref{fig:inert_0.3_traj} into dimensional quantities.  For 
instance, for the ``hoverer'' in Fig. \ref{fig:chi0_0.85_traj}, the particle is clearly 
close to the rheotactic state by $T/T_{0} = 250$.  This corresponds to an experimental 
time of ca $31 \, \mathrm{s}$.  Likewise, for the half-covered particle in Fig. 
\ref{fig:inert_0.3_traj}, the particle has turned upstream by $T/T_{0} \approx 350$, 
corresponding to an experimental time of ca $44 \, \mathrm{s}$. Furthermore, for the two 
particle designs we can evaluate dimensional values of the relaxation time scale in Eq. 
(\ref{eq:px_exp_decay}), which we denote as $\tau_{a}$.  For the hoverer, 
the dimensionless relaxation time is $\tau_{a}/T_{0} = 25.4$ (dotted line in 
Fig. \ref{fig:chi0_0.85_rheo_plots}(e)), which renders $\tau_{a} \approx 3.2 \, 
\textrm{s}$. Likewise, for the half-covered particle, the dimensionless relaxation time is 
$\tau_{a}/T_{0} = 149$ (dotted line in Fig. \ref{fig:inert_0.3_rheo_plots}(e)), which 
renders $\tau_{a} \approx 19 \, \textrm{s}$. We can compare these values for $\tau_{a}$ 
with the time scale $\tau_{r}$ for rotational diffusion in order to estimate how robust 
rheotaxis is against thermal noise. For a particle with $R = 2.5 \; \mathrm{\mu m}$ in 
water at room temperature, the time scale $\tau_{r}$ for reorientation by rotational 
diffusion is $D_{r}^{-1} = 8 \pi \mu R^{3}/k_{B} T \approx 95 \, \textrm{s}$ 
\cite{uspal14}. Therefore, the rheotactic state is expected to be robust against thermal 
noise induced rotation out of the plane $p_{x} = 0$.

Our theoretical findings are not restricted to catalytically active Janus particles. 
In order to highlight their wide range of applicability, we have also 
investigated rheotaxis for a spherical ``squirmer.'' The squirmer model captures essential 
features of the 
motion of ciliated micro-organisms \cite{drescher09} and self-propelled liquid 
droplets 
\cite{thutupalli11}.  Recently, it was shown theoretically that, for certain values of the 
leading 
order squirming mode amplitudes, a squirmer can be attracted to a planar surface and can 
move at a steady 
height and orientation \cite{ishimoto13,li14}. In the presence of shear this attraction 
can allow 
one to fulfill the two criteria we have established for the occurrence of 
rheotaxis, and 
we have shown that rheotaxis does indeed emerge. It is possible that the mechanism for 
rheotaxis outlined 
in the present study is relevant for spherical or quasi-spherical micro-organisms, such 
as 
\textit{Volvox carteri}. More speculatively, perhaps such organisms can dynamically adjust 
their 
squirming mode amplitudes to ``turn on'' and ``turn off'' rheotaxis.  However, given the 
complexity 
of these organisms -- \textit{Volvox}, for instance, is bottom-heavy and spins around a 
fixed body axis \cite{drescher09} -- 
experiments and more detailed numerical studies are needed to 
assess in 
this context the relevance of the mechanisms studied here. Finally, we note that, 
while we 
focus on active particles which can be modeled via an effective slip velocity, we 
expect that 
our theoretical findings will be applicable to other spherical active particles with an 
axisymmetric 
propulsion mechanism, provided one can justify the linear separation into the two 
subproblems considered here of an active sphere in a quiescent fluid and an inert sphere in flow.

A natural extension of our work would be to study elongated (e.g., ellipsoidal) 
active 
particles with an axisymmetric propulsion mechanism.  The classic work of Jeffery 
demonstrates that inert ellipsoidal particles in unbounded shear flow rotate in 
periodic orbits \cite{jeffery22}. More 
recently, it was shown experimentally \cite{kaya09} and numerically \cite{pozrikidis05} 
that these 
orbits are preserved (although quantitatively modified) if the particle is near, but not 
in steric 
contact with, a planar bounding surface. Theoretical and numerical studies of 
elongated active particles could shed light on how this orbital motion is transformed 
into 
the ``weather vane'' mechanism of rheotaxis with the addition of active motion. An 
interesting point of comparison would be with Ref. \cite{kantsler14}, which presents a 
mathematical 
model for the rheotaxis of a spermatozoon in steric contact with a surface. Additionally, 
we 
anticipate that ellipsoidal artificial microswimmers would be able to rheotax at higher 
shear rates 
than spherical particles, owing to the ``weather vane'' mechanism. Testing this 
expectation and quantifying the difference with spherical particles would help to guide 
the 
design and the optimization of artificial rheotactic active particles.

\acknowledgements
The authors wish to thank C. Pozrikidis for making freely available the \texttt{BEMLIB} 
library, which was used for the present numerical computations \cite{pozrikidis02}. W.E.U,
M.T., and M.N.P. acknowledge financial support from the German Science Foundation (DFG), 
grant no. TA 959/1-1.

\bibliography{janus_rheotaxis_revised}

\end{document}